\documentclass[epj]{svjour}
\usepackage{graphics}

\usepackage{amsmath}
\begin{document}
\title{Study of $\beta^- $ decays for sd model space with Shell model}
\author{}
\author{Surender Siwach$^1$\thanks{Electronic address: surender@bhu.ac.in}}
\institute{Department of Physics, Banaras Hindu University, Varanasi 221 005, Uttar Pradesh, India}
\date{\today}

\abstract{   
This paper investigates the beta decay of nuclei within the sd model space, encompassing the $0d_{3/2}$, $0d_{5/2}$, and $1s_{1/2}$ shells. We comprehensively analyze their decay characteristics, including the half-life, $logft$ value, Q value, branching ratio, B(GT) value, M(GT), and R(GT) value. These results are compared with experimental data from Z = 8-19 nuclei, employing the nuclear shell model framework. The calculations are performed using the USDB\cite{usdb} interaction, which is well-suited for nuclei with active valence nucleons in the $0d_{3/2}$, $0d_{5/2}$, and $1s_{1/2}$ shells. The study offers valuable insights into the beta decay properties of these nuclei and contributes to our understanding of the nuclear shell structure.}

\authorrunning{}
\maketitle

\maketitle
\section{Introduction \label{intro}}
Nuclear physics and astrophysical phenomena, such as supernova explosions and nucleosynthesis, heavily rely on beta decay and electron capture reactions. Neutrino capture reactions also have relevance in energetic contexts, such as supernova explosions \cite{supernova}. Beta decay provides a direct approach to measure the absolute GT transition strengths B(GT), enabling the investigation of half-lives, Q, $\beta$-values, and branching ratios in the Q-window. For studying the relative values of B(GT) strengths up to high excitation energies, charge exchange reactions like (p, n) and (3He,t) are beneficial tools. Recent experimental advancements have allowed for the comparison of GT transitions studied in charge exchange reactions and beta decays, utilizing the isospin symmetry to obtain experimental information for unstable nuclei. Through the application of these techniques, a long series of high-quality experiments have furnished new experimental information regarding the Gamow-Teller strength distribution in medium-mass nuclei.\par

In astrophysics, beta decay processes are of great significance. They play a crucial role in various astrophysical phenomena and stellar evolution. For example, beta decay reactions, such as the proton-proton chain and the CNO cycle, are responsible for stars' energy generation and nucleosynthesis process. Therefore the accurate determination of $g_A$ is essential in understanding the rates of these reactions. The effective axial-vector coupling constant $g_A$ determines the strength of the interaction between the beta particle and the nucleus during decay. It quantifies the strength of the coupling between the weak force and the particle's spin. Hence the accurate determination of $g_A$ gives us insight into various astrophysical processes.

Medium-mass neutron-rich nuclei have attracted significant attention in the field of nuclear physics, prompting numerous investigations for $^{16-24}$O [\cite{19O}-\cite{24O}]. 
In this paper, we present calculations of various quantities for neutron-rich nuclei with proton number Z = 8—17 and neutron number N $\geq$ 11, including allowed $\beta$-decay half-lives, Q values, excitation energies, $log ft$ values, branching fractions. 
These calculations are performed within sd shell model space utilizing the USDB interaction and the computational tool NuShellX \cite{nushellx}. Notably, some nuclei near the neutron drip line have not been previously studied in terms of their $\beta$-decays, making our research particularly valuable for future experimental endeavors.\par
The structure of this paper is outlined as follows: In Section 2, we provide the methodology for calculating the half-lives of beta decay. Section 3 includes information about the shell model spaces and effective interactions used in our calculations. Moving on to Section 4, we present our theoretical results and compare them with available experimental data. Finally, in Section 5, we provide a summary and draw conclusions based on our findings.


\section{Methodology}

During beta decay, the parent nuclei transition from their ground state to various excited states in the daughter nuclei (limited to those within the energy window determined by the Q-value), according to the beta-decay selection rule. The half-lives for beta decay in sd shell nuclei are calculated using the \textit{ft}-value.
$$
ft = \frac{6177}{[(g_A)^2 B(GT) + B(F) ]}
$$
Where the axial-vector coupling constant of weak interactions is denoted by $g_A (= 1.260)$. The phase-space integral that incorporates lepton kinematics is represented by f. Meanwhile, the Gamow-Teller and Fermi matrix elements are denoted by B(GT) and B(F), respectively.

The total half-life is calculated as
\begin{equation}
t_{1/2}= \left({\sum_i {\frac{1}{t_i}}}\right)^{-1}
\end{equation}
where the partial decay half-life of the daughter's state i is denoted as $t_i$. The partial half-life of allowed $\beta$ decay is expressed as:
\begin{equation}
 t_{i}= 10^{logft - logf_A}
\end{equation}
In this context, $f_A$ represents the Gamow-Teller (axial-vector) phase space vector. As $f_t$ values are significant, they are commonly presented in the form of 'log\textit{ft} values'. The log $ft$ value is defined as log$ft$ $\equiv$ log$_{10}(f_At_{i}[s])$.

The partial half-life $t_i$ is related to the total half-life $t_{1/2}$ of the allowed $\beta^-$-decay as
\begin{equation}
 t_{i}= {\frac{t_{1/2}}{b_r}}
\end{equation}
where,  ${b_r}$ is the branching ratio for the level with partial half-life ${t_i}$. The
$B(GT)$ is the Gamow-Teller matrix element
\begin{equation}
 B(GT)= \left(\frac{g_A}{g_V}\right)^2{\langle{\sigma\tau}\rangle}^2
\end{equation}
The nuclear matrix element of Eqn. (4) the Gamow-Teller operator is
\begin{equation}
 {\langle{\sigma\tau}\rangle} = \frac{{\langle {f}|| \sum_{k}{\sigma^k\tau_{\pm}^k} ||i \rangle}}{\sqrt{2J_i + 1}},
\end{equation}
where $f$ and $i$ refer to all the quantum numbers needed to specify the final and initial states, respectively, $\pm$ refers to $\beta^{\pm}$ decay,
$\tau_{\pm} = \frac{1}{2}(\tau_x + i\tau_y)$ with $\tau_+p$ = $n$, $\tau_-n$ = $p$, and $J_i$ is the total angular momentum of the initial-state. The shell model calculations enable us to present the $\beta$ decay Q value, which is determined theoretically. The theoretical $\beta$ decay Q value is expressed as:
\begin{equation}
Q = (E(SM)_i + E(C)_i) - (E(SM)_f + E(C)_f)
\end{equation}

The expression incorporates E(SM), which represents the nuclear binding energy of the interaction between the valence particles within the shell model calculation. E(C) refers to the Coulomb energy of the valence space. The subscripts i and f correspond to the parent and daughter nuclei.
\begin{table*}[htbp]
    \centering
    \caption{Comparative Analysis of Theoretical and Experimental Decay States with Calculated Shell Model Results using USDB Interaction}
    \label{tab:mytable}
    \begin{tabular}{ccccccccc}        \hline \hline
            & &  \multicolumn{2}{c}{Energy level }  &\multicolumn{2}{c}{Branching Ratio } & \multicolumn{2}{c}{$log ft$ } & B(GT)\\
        \cline{3-4} \cline{7-8}
        Parent (J$_{i}$,T$_{i})[Ref.]$ & Daughter J$_{f}$,T$_{f}$& Th. & Expt. &  Th.(\%) &  Expt.(\%) &  (Th). & (Exp.).&   (Th.)\\

        \hline
$^{19}$O,5/2,3/2\cite{19O}   &  $^{19}$F, ,$5/2_{1},1/2$   &  77   &  197.143   &  20.987   &  45.4   &  5.801   &  5.382   &  0.0062\\
   &  $3/2_{1},1/2$   &  1770   &  1554.038   &  79.0128   &  54.4   &  4.379   &  4.623   &  0.1627\\
   &  $7/2_{1},1/2$   &  4868   &  4377.7   &  0   &  0.0984   &  3.504   &  3.857   &  1.218\\
   &  $5/2_{2},1/2$   &  4969   &  4550   &  0   &  $\leq$0.001   &  7.72   &  >5.1   &  0.0001\\
   &  $3/2_{2},1/2$   &  6819   &  3908.17   &  0   &  0.0081   &  5.003   &  6.13   &  0.0386\\
$^{20}$O,0,2\cite{20O}   &  $^{20}$F,$1_{1},1$   &  1208   &  1056.848   &  97.7005   &  99.973   &  3.705   &  3.734   &  0.7674\\
   &  \hspace{0.30in}$1_{2},1$   &  3528   &  3488.41   &  2.2995   &  0.027   &  3.575   &  3.64   &  1.035\\
$^{21}$O,5/2,5/2\cite{21O}   &  $^{21}$F,$5/2_{1},3/2$   &  0   &  0   &  8.0681   &  0   &  6.361   &   *  &  0.0017\\
   &  $3/2_{1},3/2$   &  1804   &  1730.4   &  33.1716   &  37.2   &  5.236   &  5.22   &  0.0226\\
   &  $5/2_{2},3/2$   &  3499   &   *  &  29.6415   &  29.6   &  4.658   &  4.66   &  0.0854\\
   &  $3/2_{2},3/2$   &  3561   &   *  &  15.4338   &  12.3   &  4.915   &  5.06   &  0.0473\\
   &  $7/2_{1},3/2$   &  3588   &   *  &  1.5963   &   *  &  5.889   &   *  &  0.005\\
   &  $5/2_{3},3/2$   &   *  &   *  &  0.8667   &   *  &  5.868   &   *  &  0.0053\\
   &  $3/2_{3},3/2$   &   *  &   *  &  5.0825   &   *  &  4.991   &   *  &  0.0397\\
   &  $7/2_{2},3/2$   &   *  &   *  &  6.1384   &   *  &  4.884   &   *  &  0.0508\\
   &  $7/2_{3},3/2$   &   *  &   *  &  0.0012   &   *  &  8.116   &   *  &  0\\
$^{22}$O,0,3\cite{22O}   &  $^{22}$F,$1_{1},2$   &  1533   &  1627   &  18.5581   &  31   &  4.578   &  4.6   &  0.1027\\
   & \hspace{0.30in} $1_{2},2$   &  2511   &  2571.5   &  81.4419   &  68   &  3.607   &  3.8   &  0.961\\
$^{23}$O,1/2,7/2\cite{nndc}   &  $^{23}$F,$1/2_{1},5/2$   &  1991   &  2244   &  43.7428   &  45.8   &  4.497   &   *  &  0.1239\\
   & \hspace{0.30in} $3/2_{1},5/2$   &  3975   &   *  &  25.2553   &   *  &  4.337   &   *  &  0.1792\\
   & \hspace{0.30in} $3/2_{2},5/2$   &  4479   &   *  &  3.5176   &   *  &  5.079   &   *  &  0.0325\\
   & \hspace{0.30in} $1/2_{2},5/2$   &  5493   &   *  &  0.0322   &   *  &  6.867   &   *  &  0.0005\\
   & \hspace{0.30in} $3/2_{3},5/2$   &   *  &   *  &  5.7789   &   *  &  4.437   &   *  &  0.1423\\
   & \hspace{0.30in} $1/2_{3},5/2$   &   *  &   *  &  8.7683   &   *  &  4.151   &   *  &  0.275\\
   & \hspace{0.30in} $3/2_{4},5/2$   &   *  &   *  &  9.2429   &   *  &  4.108   &   *  &  0.3036\\
   & \hspace{0.30in} $3/2_{4},5/2$   &   *  &   *  &  0.8655   &   *  &  5.033   &   *  &  0.036\\
   & \hspace{0.30in} $1/2_{5},5/2$   &   *  &   *  &  2.7964   &   *  &  4.079   &   *  &  0.324\\
   & \hspace{0.30in} $1/2_{5},5/2$   &   *  &   *  &  0.001   &   *  &  8.454   &   *  &  0\\
$^{24}$O,0,4\cite{24O}   &  $^{24}$F,$1_{1},3$   &  1583   &  1831.4   &  100   &  57   &  3.799   &  4.09   &  0.6174\\
$^{20}$F,2,1\cite{20F}   &  $^{20}$Ne,$2_{1},0$   &  1747   &  1633.674   &  99.7575   &  99.9913   &  4.816   &  4.9697   &  0.0594\\
   & \hspace{0.30in}$2_{2},0$   &  7543   &  4966.51   &  0.2425   &  0.0082   &  4.109   &  7.2   &  0.3028\\
   & \hspace{0.30in}$3_{1},0$   &   *  &   *  &  0   &  0   &  5.013   &   *  &  0.0377\\
   & \hspace{0.30in}$3_{2},0$   &   *  &   *  &  0   &  0   &  4.585   &   *  &  0.1012\\
   & \hspace{0.30in}$1_{2},0$   &   *  &   *  &  0   &  0   &  3.67   &   *  &  0.8319\\
   & \hspace{0.30in}$1_{3},0$   &   *  &   *  &  0   &  0   &  4   &   *  &  0.3893\\
$^{21}$F,5/2,3/2\cite{21F}   &  $^{21}$Ne,$3/2_{1},1/2$   &  0   &  0   &  12.8035   &  9.6   &  5.427   &  5.67   &  0.0145\\
   &  \hspace{0.30in}$5/2_{1},1/2$   &  266   &  350.727   &  71.3677   &  74.1   &  4.585   &  4.65   &  0.1011\\
   &  \hspace{0.30in}$7/2_{1},1/2$   &  1757   &  1745.91   &  15.7346   &  16.1   &  4.605   &  4.72   &  0.0967\\
   &  \hspace{0.30in}$5/2_{2},1/2$   &  3718   &  3755.59   &  0.0059   &  0.003   &  6.719   &  7.11   &  0.0007\\
   &  \hspace{0.30in}$5/2_{2},1/2$   &  4627   &  4525.84   &  0.0552   &  0.043   &  4.639   &  5.02   &  0.0894\\
   &  \hspace{0.30in}$3/2_{3},1/2$   &  4914   &  4684.56   &  0.0329   &  0.077   &  4.328   &  4.5   &  0.1827\\
   &  \hspace{0.30in}$7/2_{2},1/2$   &   *  &   *  &  0.0001   &   *  &  5.284   &   *  &  0.0202\\
   &  \hspace{0.30in}$3/2_{3},1/2$   &   *  &   *  &  0   &   *  &  4.349   &   *  &  0.\\
   &  \hspace{0.30in}$7/2_{3},1/2$   &   *  &   *  &   *  &   *  &  4.863   &   *  &  0.0534\\
$^{22}$F,4,2\cite{22F}   &  $^{22}$Ne,$4_{1},1$   &  3357   &  3357.2   &  1.2189   &  3.1   &  6.666   &  6.6   &  0.0008\\
   &  \hspace{0.30in}$4_{2},1$   &  5367   &  5523.3   &  52.8062   &  53.9   &  4.606   &  4.79   &  0.0965\\
   &  \hspace{0.30in}$3_{1},1$   &  5461   &  5641.2   &  16.7565   &  16.4   &  5.082   &  5.26   &  0.0322\\
   &  \hspace{0.30in}$4_{3},1$   &  6300   &  6345.1   &  3.7067   &  7   &  5.53   &  5.34   &  0.0115\\
   &  \hspace{0.30in} $3_{2},1$   &   *  &   *  &  0.7471   &   *  &  6.17   &   *  &  0.0026\\
   &  \hspace{0.30in}$5_{1},1$   &  7294   &  7423.4   &  18.7559   &  8.7   &  4.551   &  4.7   &  0.1093\\
   &  \hspace{0.30in}$3_{3},1$   &   *  &   *  &  4.8036   &   *  &  5.992   &   *  &  0.004\\
   &  \hspace{0.30in}$5_{2},1$   &   *  &   *  &  0.7451   &   *  &  4.719   &   *  &  0.0742\\
   &  \hspace{0.30in}$5_{3},1$   &   *  &   *  &   *  &   *  &  5.299   &   *  &  0.019\\
$^{23}$F,5/2,5/2\cite{23F}   &  $^{23}$Ne,$5/2_{1},3/2$   &  0   &  0   &  34.160   &  30   &  5.417   &  5.7   &  0.0149\\
   &  \hspace{0.30in}$7/2_{1},3/2$   &  1809   &  1701   &  0.685   &   *  &  6.629   &   *  &  0.0158\\
   &  \hspace{0.30in}$3/2_{1},3/2$   &  1875   &  1822   &  5.092   &  11   &  5.735   &  5.6   &  0.023\\
   &  \hspace{0.30in}$5/2_{2},3/2$   &  2330   &  2314   &  1.273   &  5.6   &  6.192   &  5.8   &  0.0255\\
   &  \hspace{0.30in}$3/2_{2},3/2$   &  3330   &  3431   &  17.604   &  12   &  4.695   &  5   &  0.104\\
   &  \hspace{0.30in}$5/2_{2},3/2$   &  3722   &  3830.9   &  27.879   &  22   &  4.337   &  4.6   &  0.283\\
   &  \hspace{0.30in}$7/2_{2},3/2$   &  3776   &  3830   &  1.577   &   *  &  5.562   &   *  &  0.2937\\
   &  \hspace{0.30in}$3/2_{3},3/2$   &   *  &   *  &  0.5801   &   *  &  5.971   &   *  &  0.2979\\
   &  \hspace{0.30in}$7/2_{3},3/2$   &  4358   &  4435.9   &  10.49   &  9   &  4.477   &  4.7   &  0.4276\\
\hline

    \end{tabular}
    \begin{flushleft}
        Please note that '*' indicates unresolved or unavailable data.
    \end{flushleft}
    
\end{table*}


\addtocounter{table}{-1}

\begin{table*}
  \begin{center}
    \leavevmode
    \caption{{\em Continuation.\/}}   

    \label{tab:mgt_comp} 
   \begin{tabular}{ccccccccc}        \hline \hline
            & &  \multicolumn{2}{c}{Energy level }  &\multicolumn{2}{c}{Branching Ratio } & \multicolumn{2}{c}{$log ft$ } & B(GT)\\
        \cline{3-4} \cline{7-8}
        Parent (J$_{i}$,T$_{i})[Ref.]$ & Daughter J$_{f}$,T$_{f}$& Th. & Expt. &  Th.(\%) &  Expt.(\%) &  (Th). & (Exp.).&   (Th.)\\

         \hline  
           &  \hspace{0.30in}$5/2_{3},3/2$   &   *  &   *  &  0.0062   &   *  &  7.387   &   *  &  0.4277\\
   &  \hspace{0.30in}$3/2_{4},3/2$   &   *  &   *  &  0.0739   &   *  &  6.31   &   *  &  0.4296\\
   &  \hspace{0.30in}$5/2_{4},3/2$   &   *  &   *  &  0.1132   &   *  &  5.799   &   *  &  0.4358\\
   &  \hspace{0.30in}$7/2_{3},3/2$   &   *  &   *  &  0.1492   &   *  &  5.63   &   *  &  0.4449\\
   &  \hspace{0.30in}$3/2_{4},3/2$   &   *  &   *  &  0.1124   &   *  &  5.713   &   *  &  0.4525\\
   &  \hspace{0.30in}$7/2_{4},3/2$   &   *  &   *  &  0.2089   &   *  &  5.349   &   *  &  0.4699\\
$^{24}$F,3,3\cite{24F}   &  $^{24}$Ne,$2_{1},2$   &  2111   &  1981.6   &  22.8166   &   *  &  5.544   &   *  &  0.0111\\
   &  \hspace{0.30in}$2_{2},2$   &   *  &   *  &  21.922   &   *  &  5.293   &   *  &  0.0198\\
   &  \hspace{0.30in}$4_{1},2$   &   *  &   *  &  44.8041   &   *  &  4.941   &   *  &  0.0446\\
   &  \hspace{0.30in}$3_{1},2$   &   *  &   *  &  4.4049   &   *  &  5.796   &   *  &  0.0062\\
   &  \hspace{0.30in}$3_{2},2$   &   *  &   *  &  0.2183   &   *  &  6.978   &   *  &  0.0004\\
   &  \hspace{0.30in}$4_{2},2$   &   *  &   *  &  5.8341   &   *  &  5.499   &   *  &  0.0123\\
$^{25}$F,5/2,7/2\cite{25F}   &  $^{25}$Ne,$3/2_{1},5/2$   &  2043   &  1702   &  19.6223   &  18   &  4.983   &  5.2   &  0.0404\\
   &  \hspace{0.30in}$3/2_{2},5/2$   &  3169   &   *  &  3.2086   &   *  &  5.588   &   *  &  0.01\\
   &  \hspace{0.30in}$7/2_{1},5/2$   &  3770   &   *  &  30.4252   &   *  &  4.507   &   *  &  0.1211\\
   &  \hspace{0.30in}$7/2_{2},5/2$   &  4451   &   *  &  30.4252   &   *  &  4.507   &   *  &  0.1211\\
   &  \hspace{0.30in}$7/2_{3},5/2$   &  5060   &   *  &  0.327   &   *  &  6.351   &   *  &  0.0017\\
   &  \hspace{0.30in}$7/2_{4},5/2$   &  6047   &   *  &  0.327   &   *  &  6.351   &   *  &  0.0017\\
   &  \hspace{0.30in}$3/2_{2},5/2$   &  6160   &   *  &  2.3347   &   *  &  5.463   &   *  &  0.0134\\
   &  \hspace{0.30in}$7/2_{5},5/2$   &   *  &   *  &  0.761   &   *  &  5.867   &   *  &  0.0053\\
   &  \hspace{0.30in}$7/2_{6},5/2$   &   *  &   *  &  0.761   &   *  &  5.867   &   *  &  0.0053\\
   &  \hspace{0.30in}$7/2_{7},5/2$   &   *  &   *  &  4.6816   &   *  &  4.872   &   *  &  0.0522\\
   &  \hspace{0.30in}$7/2_{8},5/2$   &   *  &   *  &  4.6816   &   *  &  4.872   &   *  &  0.0522\\
   &  \hspace{0.30in}$3/2_{3},5/2$   &   *  &   *  &  2.4449   &   *  &  5.13   &   *  &  0.0289\\
$^{26}$F,4,4\cite{26F}   &  $^{26}$Ne,$4_{1},3$   &   *  &   *  &  73.6897   &   *  &  4.383   &   *  &  0.1609\\
   &  \hspace{0.30in}$3_{1},3$   &   *  &   *  &  3.3661   &   *  &  5.499   &   *  &  0.0123\\
   &  \hspace{0.30in}$4_{2},3$   &   *  &   *  &  13.3773   &   *  &  4.887   &   *  &  0.0505\\
   &  \hspace{0.30in}$3_{2},3$   &   *  &   *  &  1.9842   &   *  &  5.627   &   *  &  0.002\\
   &  \hspace{0.30in}$5_{1},3$   &   *  &   *  &  1.7723   &   *  &  5.43   &   *  &  0.0145\\
   &  \hspace{0.30in}$5_{2},3$   &   *  &   *  &  5.8103   &   *  &  4.853   &   *  &  0.0546\\
$^{26}$F,1,4   &  $^{26}$Ne,$0_{1},3$   &  0   &   *  &  35.2724   &   *  &  4.537   &   *  &  0.1129\\
   &  \hspace{0.30in}$2_{1},3$   &  2063   &   *  &  48.2307   &   *  &  4.15   &   *  &  0.2755\\
   &  \hspace{0.30in}$2_{2},3$   &  3767   &   *  &  10.7805   &   *  &  4.568   &   *  &  0.1053\\
   &  \hspace{0.30in}$0_{2},3$   &  4636   &   *  &  4.2856   &   *  &  4.839   &   *  &  0.0564\\
   &  \hspace{0.30in}$1_{1},3$   &  4725   &   *  &  0.175   &   *  &  6.214   &   *  &  0.0024\\
   &  \hspace{0.30in}$1_{2},3$   &  6592   &   *  &  1.2559   &   *  &  5.048   &   *  &  0.0348\\
$^{23}$Ne,5/2,3/2\cite{23Ne}   &  $^{23}$Na,$3/2_{1},1/2$   &  0   &  0   &  57.9673   &  67   &  5.469   &  5.27   &  0.0132\\
   &  \hspace{0.30in}$5/2_{1},1/2$   &  440.3   &  399   &  41.2246   &  31.9   &  5.429   &  5.38   &  0.0145\\
   &  \hspace{0.30in}$7/2_{1},1/2$   &  2168   &  2076   &  0.3014   &  1.1   &  6.44   &  5.82   &  0.0014\\
   &  \hspace{0.30in}$3/2_{2},1/2$   &  2981.8   &  2723   &  0.5064   &  0.065   &  5.685   &  6.13   &  0.008\\
   &  \hspace{0.30in}$5/2_{2},1/2$   &  3749   &   *  &  0.0003   &   *  &  7.216   &   *  &  0.0002\\
$^{24}$Ne,0,2\cite{24Ne}   &  $^{24}$Na,$1_{1},1$   &  540   &  472.2073   &  84.8747   &  92.1   &  4.278   &  4.364   &  0.2052\\
   &  \hspace{0.30in}$1_{2},1$   &  1324   &  1346.64   &  15.1253   &  7.9   &  4.323   &  4.4   &  0.1851\\
$^{25}$Ne,1/2,5/2\cite{25Ne}   &  $^{25}$Na,$3/2_{1},3/2$   &  114   &  89.53   &  69.0255   &  76.6   &  4.341   &  4.41   &  0.1776\\
   &  \hspace{0.30in}$1/2_{1},3/2$   &  1069.32   &  966   &  24.4595   &  19.5   &  4.54   &  4.7   &  0.1123\\
   &  \hspace{0.30in}$3/2_{2},3/2$   &  1981   &  2202   &  4.1392   &  2.1   &  4.967   &  5.26   &  0.042\\
   &  \hspace{0.30in}$3/2_{3},3/2$   &  3525   &  3687   &  1.1953   &  1.2   &  4.845   &  4.82   &  0.0556\\
   &  \hspace{0.30in}$1/2_{2},3/2$   &   *  &   *  &  1.1402   &   *  &  4.542   &   *  &  0.1116\\
   &  \hspace{0.30in}$1/2_{3},3/2$   &  4289   &  4119   &  0.0402   &  0.53   &  4.933   &  4.82   &  0.0454\\
$^{26}$Ne,0,3\cite{26Ne}   &  $^{26}$Na,$1_{1},2$   &  4   &  91.6   &  83.4445   &  91.6   &  3.7   &  3.87   &  0.776\\
   &  \hspace{0.30in}$1_{2},2$   &  1281   &  1511   &  12.4313   &  4.2   &  4.14   &  4.8   &  0.2818\\
   &  \hspace{0.30in}$1_{3},2$   &  2721   &  2450   &  3.9369   &  1.9   &  4.211   &  4.7   &  0.2393\\
   &  \hspace{0.30in}$1_{4},2$   &   *  &   *  &  0.1874   &   *  &  5.439   &   *  &  0.0142\\
$^{27}$Ne,3/2,7/2\cite{27Ne}   &  $^{27}$Na,$5/2_{1},5/2$   &  0   &  0   &  64.484   &  59.5   &  4.213   &  4.4   &  0.2381\\
   &  \hspace{0.30in}$3/2_{1},5/2$   &  28   &  63   &  14.9302   &  4.2   &  4.844   &  5.54   &  0.0557\\
   &  \hspace{0.30in}$1/2_{1},5/2$   &  1661   &  1728.1   &  0.4334   &  3.4   &  6.092   &  5.33   &  0.032\\
   &  \hspace{0.30in}$5/2_{2},5/2$   &  2.943   &   *  &  6.9833   &   *  &  4.626   &   *  &  0.0921\\
   &  \hspace{0.30in}$3/2_{2},5/2$   &  3218   &   *  &  10.2101   &   *  &  4.401   &   *  &  0.1545\\
   &  \hspace{0.30in}$1/2_{2},5/2$   &  3.859   &   *  &  0.6278   &   *  &  5.467   &   *  &  0.0133\\
   &  \hspace{0.30in}$5/2_{3},5/2$   &  3867   &   *  &  0.3812   &   *  &  5.681   &   *  &  0.0081\\
   &  \hspace{0.30in}$3/2_{3},5/2$   &  3899   &   *  &  0.5428   &   *  &  5.52   &   *  &  0.0117\\
   &  \hspace{0.30in}$1/2_{3},5/2$   &  4.71   &   *  &  1.4067   &   *  &  4.906   &   *  &  0.0484\\

                                      \hline
    \end{tabular}
    
  \end{center}
\end{table*}

\addtocounter{table}{-1}

\begin{table*}
  \begin{center}
    \leavevmode
    \caption{{\em Continuation.\/}}   

    \label{tab:mgt_comp} 
   \begin{tabular}{ccccccccc}        \hline \hline
            & &  \multicolumn{2}{c}{Energy level }  &\multicolumn{2}{c}{Branching Ratio } & \multicolumn{2}{c}{$log ft$ } & B(GT)\\
        \cline{3-4} \cline{7-8}
        Parent (J$_{i}$,T$_{i})[Ref.]$ & Daughter J$_{f}$,T$_{f}$& Th. & Expt. &  Th.(\%) &  Expt.(\%) &  (Th). & (Exp.).&   (Th.)\\

         \hline  
$^{28}$Ne,0,4\cite{28Ne}   &  $^{28}$Na,$1_{1},3$   &  117   &  0   &  72.7406   &  55   &  3.643   &  4.2   &  0.8842\\
   &  \hspace{0.30in}$1_{2},3$   &  932   &  1932.2   &  14.7253   &  1.7   &  3.967   &  5.3   &  0.4199\\
   &  \hspace{0.30in}$1_{3},3$   &  2686   &  2118.4   &  1.9715   &  20.1   &  4.725   &  4.2   &  0.0733\\
   &  \hspace{0.30in}$1_{4},3$   &  3144   &  2714.3   &  2.9957   &  8.5   &  4.444   &  4.5   &  0.14\\
   &  \hspace{0.30in}$1_{5},3$   &  4352   &  3286.4   &  7.4902   &  1.3   &  3.759   &  5.2   &  0.6772\\
   &  \hspace{0.30in}$1_{6},3$   &  4481   &  3512.5   &  0.0766   &  0.9   &  5.716   &  5.3   &  0.0075\\
$^{29}$Ne,3/2,4.5\cite{29Ne}   &  $^{29}$Na,$5/2_{1},7/2$   &  0   &  72   &  24.0716   &  33   &  4.521   &   *  &  0.1173\\
   &  \hspace{0.30in}$5/2_{2},7/2$   &  2943   &   *  &  24.0716   &   *  &  4.521   &   *  &  0.1173\\
   &  \hspace{0.30in}$1/2_{1},7/2$   &  1661   &   *  &  1.5023   &   *  &  5.419   &   *  &  0.0148\\
   &  \hspace{0.30in}$5/2_{3},7/2$   &  3867   &   *  &  21.6126   &   *  &  4.068   &   *  &  0.3329\\
   &  \hspace{0.30in}$5/2_{4},7/2$   &   *  &   *  &  21.6126   &   *  &  4.068   &   *  &  0.3329\\
   &  \hspace{0.30in}$1/2_{2},7/2$   &  3859   &   *  &  3394   &   *  &  5.852   &   *  &  0.0055\\
   &  \hspace{0.30in}$1/2_{3},7/2$   &  4710   &   *  &  5.0929   &   *  &  4.588   &   *  &  0.1004\\
   &  \hspace{0.30in}$5/2_{5},7/2$   &   *  &   *  &  0.8485   &   *  &  5.344   &   *  &  0.0176\\
   &  \hspace{0.30in}$5/2_{6},7/2$   &   *  &   *  &  0.8485   &   *  &  5.344   &   *  &  0.0176\\
$^{30}$Ne,0,5\cite{30Ne}   &  $^{30}$Na,$1_{1},4$   &  248   &  150.62   &  60.3304   &  63   &  3.71   &  4.04   &  0.7586\\
   &  \hspace{0.30in}$1_{2},4$   &  2767   &  926   &  23.1876   &  7.7   &  3.731   &  4.84   &  0.7233\\
   &  \hspace{0.30in}$1_{3},4$   &  3800   &  2113.6   &  16.482   &  14   &  3.693   &  4.39   &  0.789\\
$^{24}$Na,4,0\cite{24Na}   &  $^{24}$Mg,$4_{1},0$   &  4372   &  4122.844   &  76.280   &  99.867   &  5.693   &  6.12   &  0.0079\\
   &  \hspace{0.30in}$3_{1},0$   &  5070   &  5235.21   &  19.7423   &  0.07   &  6.011   &  6.66   &  0.0038\\
   &  \hspace{0.30in}$4_{2},0$   &  5883   &   *  &  3.9619   &   *  &  6.345   &   *  &  0.0018\\
   &  \hspace{0.30in}$5_{1},0$   &   *  &   *  &  0.0148   &   *  &  7.534   &   *  &  0.0001\\
   &  \hspace{0.30in}$3_{2},0$   &   *  &   *  &  0.0005   &   *  &  6.159   &   *  &  0.0027\\
   &  \hspace{0.30in}$5_{2},0$   &   *  &   *  &  0   &   *  &  4.249   &   *  &  0.2191\\
$^{25}$Na,5/2,3/2\cite{25Na}   &  $^{25}$Mg,$5/2_{1},1/2$   &  0   &  0   &  68.0102   &  62.5   &  5.075   &  5.26   &  0.0327\\
   &  \hspace{0.30in}$3/2_{1},1/2$   &  1097   &  974.749   &  24.5795   &  27.46   &  4.869   &  5.04   &  0.0526\\
   &  \hspace{0.30in}$7/2_{1},1/2$   &  1721   &  1611.772   &  6.6518   &  9.48   &  4.953   &  5.03   &  0.0433\\
   &  \hspace{0.30in}$5/2_{2},1/2$   &  1995   &  1964.62   &  0.3719   &  0.44   &  5.95   &  6.04   &  0.0044\\
   &  \hspace{0.30in}$3/2_{2},1/2$   &  2811   &  2801.46   &  0.385   &  0.247   &  4.899   &  5.25   &  0.049\\
   &  \hspace{0.30in}$7/2_{2},1/2$   &  2901   &   *  &  0.0015   &   *  &  7.157   &   *  &  0.0003\\
$^{26}$Na,3,2\cite{26Na}   &  $^{26}$Mg,$2_{1},1$   &  1897   &  1808.81   &  76.169   &  87.8   &  4.575   &  4.71   &  0.1036\\
   &  \hspace{0.30in}$2_{2},1$   &  3007   &  2938.26   &  0.5895   &  0.05   &  6.427   &  7.6   &  0.0015\\
   &  \hspace{0.30in}$3_{1},1$   &  3882   &  3941.48   &  1.108   &  1.31   &  5.924   &  5.87   &  0.0046\\
   &  \hspace{0.30in}$3_{2},1$   &  4317   &  4350.02   &  8.2572   &  3.17   &  4.929   &  5.33   &  0.0459\\
   &  \hspace{0.30in}$4_{1},1$   &  4365   &  4319.17   &  0.8785   &  0.493   &  5.888   &  6.15   &  0.005\\
   &  \hspace{0.30in}$2_{3},1$   &  4449   &  4332.02   &  2.5027   &  1.65   &  5.408   &  5.62   &  0.0152\\
   &  \hspace{0.30in}$2_{4},1$   &  4882   &  4834.92   &  3.3914   &  2.378   &  5.143   &  5.25   &  0.028\\
   &  \hspace{0.30in}$4_{2},1$   &  4939   &   *  &  0.0487   &   *  &  6.968   &   *  &  0.0004\\
   &  \hspace{0.30in}$2_{5},1$   &  5385   &  5291.65   &  0.0186   &  0.0129   &  7.239   &  7.31   &  0.0002\\
   &  \hspace{0.30in}$4_{3},1$   &   *  &   *  &  0.2617   &   *  &  6.043   &   *  &  0.0035\\
   &  \hspace{0.30in}$4_{4},1$   &   *  &   *  &  1.99   &   *  &  5.028   &   *  &  0.0365\\
   &  \hspace{0.30in}$3_{2},1$   &  6179   &  6125   &  4.2126   &  1.72   &  4.593   &  4.74   &  0.0994\\
   &  \hspace{0.30in}$2_{6},1$   &   *  &   *  &  0.137   &   *  &  5.876   &   *  &  0.0052\\
   &  \hspace{0.30in}$4_{5},1$   &   *  &   *  &  0.129   &   *  &  5.879   &   *  &  0.0051\\
   &  \hspace{0.30in}$3_{3},1$   &   *  &   *  &  0.2031   &   *  &  5.419   &   *  &  0.0148\\
   &  \hspace{0.30in}$4_{5},1$   &   *  &   *  &  0.0093   &   *  &  6.691   &   *  &  0.0008\\
   &  \hspace{0.30in}$3_{4},1$   &   *  &   *  &  0.0792   &   *  &  5.619   &   *  &  0.0094\\
   &  \hspace{0.30in}$3_{5},1$   &   *  &   *  &  0.0137   &   *  &  6.072   &   *  &  0.0033\\
$^{27}$Na,5/2,5/2\cite{27Na}   &  $^{27}$Mg,$3/2_{1},3/2$   &  993   &  984.69   &  83.1719   &  85.8   &  4.084   &  4.3   &  0.3204\\
   &  \hspace{0.30in}$5/2_{1},3/2$   &  1673   &  1698.06   &  6.8371   &  11.3   &  5.01   &  4.99   &  0.038\\
   &  \hspace{0.30in}$5/2_{2},3/2$   &  1903   &  1940.06   &  1.3168   &  0.5   &  5.668   &  6.3   &  0.0083\\
   &  \hspace{0.30in}$7/2_{1},3/2$   &  3084   &  3109.5   &  0.6022   &  0.5   &  5.69   &  5.91   &  0.0079\\
   &  \hspace{0.30in}$3/2_{2},3/2$   &  3459   &  3490.9   &  0.3861   &  0.52   &  5.771   &  5.76   &  0.0066\\
   &  \hspace{0.30in}$7/2_{2},3/2$   &  3465   &  3427.1   &  0.956   &  0.74   &  5.376   &  5.63   &  0.0164\\
   &  \hspace{0.30in}$3/2_{3},3/2$   &  3602   &  4150   &  0.8395   &  0.026   &  5.39   &  6.81   &  0.0159\\
   &  \hspace{0.30in}$5/2_{3},3/2$   &  4113   &   *  &  0.3604   &   *  &  5.591   &   *  &  0.01\\
   &  \hspace{0.30in}$5/2_{4},3/2$   &   *  &   *  &  0.7471   &   *  &  5.187   &   *  &  0.0253\\
   &  \hspace{0.30in}$7/2_{2},3/2$   &   *  &   *  &  0.5299   &   *  &  5.21   &   *  &  0.024\\
   &  \hspace{0.30in}$5/2_{5},3/2$   &   *  &   *  &  1.6929   &   *  &  4.62   &   *  &  0.0933\\
   &  \hspace{0.30in}$3/2_{4},3/2$   &   *  &   *  &  0.0019   &   *  &  7.505   &   *  &  0.0001\\
   &  \hspace{0.30in}$7/2_{3},3/2$   &   *  &   *  &  0.8669   &   *  &  4.837   &   *  &  0.0566\\
   &  \hspace{0.30in}$5/2_{6},3/2$   &   *  &   *  &  0.505   &   *  &  5.059   &   *  &  0.034\\
   &  \hspace{0.30in}$3/2_{5},3/2$   &   *  &   *  &  0.0304   &   *  &  6.08   &   *  &  0.0032\\
   &  \hspace{0.30in}$7/2_{4},3/2$   &   *  &   *  &  0.2598   &   *  &  5.141   &   *  &  0.0281\\
   &  \hspace{0.30in}$5/2_{7},3/2$   &   *  &   *  &  0.2747   &   *  &  5.093   &   *  &  0.0314\\

                                      \hline
    \end{tabular}
  \end{center}
\end{table*}
\addtocounter{table}{-1}

\begin{table*}
  \begin{center}
    \leavevmode
    \caption{{\em Continuation.\/}}   

    \label{tab:mgt_comp} 
   \begin{tabular}{ccccccccc}        \hline \hline
            & &  \multicolumn{2}{c}{Energy level }  &\multicolumn{2}{c}{Branching Ratio } & \multicolumn{2}{c}{$log ft$ } & B(GT)\\
        \cline{3-4} \cline{7-8}
        Parent (J$_{i}$,T$_{i})[Ref.]$ & Daughter J$_{f}$,T$_{f}$& Th. & Expt. &  Th.(\%) &  Expt.(\%) &  (Th). & (Exp.).&   (Th.)\\

         \hline $^{28}$Na,1,3\cite{28Na}   &  $^{28}$Mg,$0_{1},2$   &  0   &  0   &  59.6477   &  60.4   &  4.419   &  4.6   &  0.1483\\
   &  \hspace{0.30in}$2_{1},2$   &  1518   &  1473.55   &  4.2164   &  11   &  5.332   &  5.1   &  0.0181\\
   &  \hspace{0.30in}$0_{2},2$   &  4007   &  3862.15   &  21.1598   &  20.1   &  4.172   &  4.42   &  0.2616\\
   &  \hspace{0.30in}$2_{2},2$   &  4543   &  4554.6   &  2.3639   &  1   &  5.011   &  5.6   &  0.0379\\
   &  \hspace{0.30in}$1_{1},2$   &  4663   &  4561   &  3.5429   &  3.2   &  4.809   &  5.1   &  0.0604\\
   &  \hspace{0.30in}$2_{3},2$   &  4794   &  4878.7   &  0.0136   &  0.2   &  7.197   &  6.2   &  0.0002\\
   &  \hspace{0.30in}$1_{2},2$   &  5.518   &  5270.1   &  2.1012   &  1.5   &  4.84   &  5.2   &  0.0563\\
   &  \hspace{0.30in}$2_{4},2$   &  5567   &  5470.2   &  0.1946   &  $\leq$0.1   &  5.862   &  >6.4   &  0.054\\
   &  \hspace{0.30in}$2_{5},2$   &  6069   &  5916.9   &  1.3337   &  0.3   &  4.9   &  5.8   &  0.0489\\
   &  \hspace{0.30in}$0_{3},2$   &  6592   &  6545   &  0.3256   &  0.2   &  5.375   &  5.8   &  0.0164\\
   &  \hspace{0.30in}$2_{6},2$   &  6947   &  7200.9   &  0.2251   &  0.5   &  5.435   &  5.2   &  0.0143\\
   &  \hspace{0.30in}$1_{3},2$   &  7054   &  7461.8   &  0.6703   &  1.4   &  4.931   &  4.7   &  0.0457\\
   &  \hspace{0.30in}$0_{4},2$   &   *  &   *  &  0.4852   &   *  &  5.05   &   *  &  0.0347\\
   &  \hspace{0.30in}$1_{4},2$   &   *  &   *  &  0.139   &   *  &  5.49   &   *  &  0.0126\\
   &  \hspace{0.30in}$2_{7},2$   &   *  &   *  &  0.1039   &   *  &  5.576   &   *  &  0.0103\\
   &  \hspace{0.30in}$2_{8},2$   &   *  &   *  &  2.7145   &   *  &  4.136   &   *  &  0.2843\\
   &  \hspace{0.30in}$0_{5},2$   &   *  &   *  &  0.1972   &   *  &  5.155   &   *  &  0.0272\\
   &  \hspace{0.30in}$1_{5},2$   &   *  &   *  &  0.503   &   *  &  4.566   &   *  &  0.1056\\
   &  \hspace{0.30in}$1_{6},2$   &   *  &   *  &  0.0062   &   *  &  6.299   &   *  &  0.002\\
   &  \hspace{0.30in}$0_{6},2$   &   *  &   *  &  0.0001   &   *  &  7.951   &   *  &  0\\
   &  \hspace{0.30in}$1_{7},2$   &   *  &   *  &  0.0377   &   *  &  5.33   &   *  &  0.0182\\
   &  \hspace{0.30in}$1_{8},2$   &   *  &   *  &  0.0017   &   *  &  6.509   &   *  &  0.0012\\
   &  \hspace{0.30in}$0_{7},2$   &   *  &   *  &  0.0032   &   *  &  6.235   &   *  &  0.0023\\
   &  \hspace{0.30in}$0_{8},2$   &   *  &   *  &  0.0135   &   *  &  5.516   &   *  &  0.0119\\
$^{29}$Na,3/2,7/2\cite{29Na}   &  $^{29}$Mg,$3/2_{1},5/2$   &  0   &  0   &  35.1635   &  24   &  4.602   &  5.06   &  0.0974\\
   &  \hspace{0.30in}$1/2_{1},5/2$   &  45   &   *  &  11.581   &   *  &  5.077   &   *  &  0.0326\\
   &  \hspace{0.30in}$5/2_{2},5/2$   &  1594   &   *  &  0.1664   &   *  &  6.661   &   *  &  0.008\\
   &  \hspace{0.30in}$3/2_{2},5/2$   &  2330   &   *  &  0.5719   &   *  &  5.991   &   *  &  0.004\\
   &  \hspace{0.30in}$1/2_{2},5/2$   &  2627   &   *  &  28.3018   &   *  &  4.239   &   *  &  0.2242\\
   &  \hspace{0.30in}$5/2_{3},5/2$   &  3152   &   *  &  4.4167   &   *  &  4.942   &   *  &  0.0445\\
   &  \hspace{0.30in}$3/2_{3},5/2$   &  3502   &   *  &  10.6317   &   *  &  4.488   &   *  &  0.1265\\
   &  \hspace{0.30in}$5/2_{4},5/2$   &  3568   &   *  &  3.8083   &   *  &  4.92   &   *  &  0.0468\\
   &  \hspace{0.30in}$5/2_{5},5/2$   &  4264   &   *  &  0.3995   &   *  &  5.746   &   *  &  0.007\\
   &  \hspace{0.30in}$3/2_{4},5/2$   &  4891   &   *  &  4.8335   &   *  &  7.8   &   *  &  0.001\\
   &  \hspace{0.30in}$1/2_{3},5/2$   &  5433   &   *  &  0.1232   &   *  &  4.38   &   *  &  0.1623\\
   &  \hspace{0.30in}$1/2_{4},5/2$   &  6093   &   *  &   *  &   *  &  5.795   &   *  &  0.0062\\
$^{30}$Na,2,4\cite{30Na}   &  $^{30}$Mg,$2_{1},3$   &  1591   &  1482.53   &  9.4971   &  9.5   &  5.269   &  5.86   &  0.021\\
   &  \hspace{0.30in}$2_{2},3$   &  3433   &  2467.8   &  10.3898   &  3.8   &  4.996   &  6.12   &  0.0393\\
   &  \hspace{0.30in}$3_{1},3$   &  4661   &  3461.4   &  0.6652   &   *  &  6.018   &   *  &  0.0037\\
   &  \hspace{0.30in}$2_{3},3$   &  4789   &  3541.5   &  28.2814   &   *  &  4.371   &   *  &  0.1656\\
   &  \hspace{0.30in}$2_{4},3$   &  5147   &  4414.9   &  1.2435   &   *  &  5.674   &   *  &  0.0082\\
   &  \hspace{0.30in}$1_{1},3$   &  5166   &   *  &  7.4875   &   *  &  4.892   &   *  &  0.0499\\
   &  \hspace{0.30in}$1_{2},3$   &   *  &   *  &  8.878   &   *  &  4.736   &   *  &  0.0715\\
   &  \hspace{0.30in}$3_{2},3$   &   *  &   *  &  5.8731   &   *  &  4.772   &   *  &  0.0657\\
   &  \hspace{0.30in}$2_{5},3$   &   *  &   *  &  6.0893   &   *  &  4.726   &   *  &  0.0732\\
   &  \hspace{0.30in}$2_{6},3$   &   *  &   *  &  0.0782   &   *  &  6.602   &   *  &  0.001\\
   &  \hspace{0.30in}$3_{3},3$   &   *  &   *  &  5.1763   &   *  &  4.75   &   *  &  0.0692\\
   &  \hspace{0.30in}$2_{7},3$   &   *  &   *  &  0.6439   &   *  &  5.612   &   *  &  0.0095\\
   &  \hspace{0.30in}$3_{4},3$   &   *  &   *  &  1.1474   &   *  &  5.336   &   *  &  0.0179\\
   &  \hspace{0.30in}$1_{3},3$   &   *  &   *  &  3.2149   &   *  &  4.87   &   *  &  0.0524\\
   &  \hspace{0.30in}$2_{4},3$   &   *  &   *  &  0.4182   &   *  &  5.701   &   *  &  0.077\\
   &  \hspace{0.30in}$2_{5},3$   &   *  &   *  &  2.7423   &   *  &  4.823   &   *  &  0.0586\\
   &  \hspace{0.30in}$3_{5},3$   &   *  &   *  &  0.0097   &   *  &  7.244   &   *  &  0.0002\\
   &  \hspace{0.30in}$3_{6},3$   &   *  &   *  &  2.5133   &   *  &  4.769   &   *  &  0.0662\\
   &  \hspace{0.30in}$1_{4},3$   &   *  &   *  &  0.1351   &   *  &  6.018   &   *  &  0.0037\\
   &  \hspace{0.30in}$3_{7},3$   &   *  &   *  &  1.793   &   *  &  4.851   &   *  &  0.0548\\
   &  \hspace{0.30in}$1_{5},3$   &   *  &   *  &  0.0078   &   *  &  7.194   &   *  &  0.0002\\
   &  \hspace{0.30in}$3_{8},3$   &   *  &   *  &  0.0809   &   *  &  6.132   &   *  &  0.0029\\
   &  \hspace{0.30in}$1_{6},3$   &   *  &   *  &  0.0134   &   *  &  6.893   &   *  &  0.0005\\
$^{31}$Na,3/2,4.5\cite{31Na}   &  $^{31}$Mg,$3/2_{1},7/2$   &  0   &  49.93   &  9.69   &   *  &  4.948   &   *  &  0.0439\\
   &  \hspace{0.30in}$5/2_{1},7/2$   &   *  &  944   &  9.365   &   *  &  4.765   &   *  &  0.0668\\
  
                                      \hline
    \end{tabular}
  \end{center}
\end{table*}
\addtocounter{table}{-1}

\begin{table*}
  \begin{center}
    \leavevmode
    \caption{{\em Continuation.\/}}   

    \label{tab:mgt_comp} 
   \begin{tabular}{ccccccccc}        \hline \hline
            & &  \multicolumn{2}{c}{Energy level }  &\multicolumn{2}{c}{Branching Ratio } & \multicolumn{2}{c}{$log ft$ } & B(GT)\\
        \cline{3-4} \cline{7-8}
        Parent (J$_{i}$,T$_{i})[Ref.]$ & Daughter J$_{f}$,T$_{f}$& Th. & Expt. &  Th.(\%) &  Expt.(\%) &  (Th). & (Exp.).&   (Th.)\\

         \hline$^{27}$Mg,1/2,3/2\cite{27Mg}   &  $^{27}$Al,$1/2_{1},1/2$   &  882   &  843.76   &  70.6176   &  70.94   &  4.54   &  4.7297   &  0.1122\\
   &  \hspace{0.30in}$3/2_{1},1/2$   &  1014   &  1014.56   &  29.3824   &  29.06   &  4.786   &  4.934   &  0.0637\\
$^{28}$Mg,0,2\cite{nndc}   &  $^{28}$Al,$1_{1},1$   &  1198   &  1372.85   &  77.0918   &  94.8   &  4.201   &  4.453   &  0.2449\\
   &  \hspace{0.30in}$1_{2},1$   &  1500   &  1620.1   &  22.9082   &  4.9   &  4.402   &  4.57   &  0.1541\\
$^{29}$Mg,3/2,5/2\cite{29Al}   &  $^{29}$Al,$5/2_{1},3/2$   &  0   &  0   &  40.8769   &  27   &  4.867   &  5.32   &  0.0528\\
   &  \hspace{0.30in}$1/2_{1},3/2$   &  1214   &  1398.05   &  4.3222   &  7   &  5.491   &  5.49   &  0.0126\\
   &  \hspace{0.30in}$3/2_{1},3/2$   &  2076   &  2224.1   &  12.0053   &  21   &  4.756   &  4.73   &  0.0682\\
   &  \hspace{0.30in}$3/2_{2},3/2$   &  2704   &  2865.6   &  7.1292   &  7.8   &  4.742   &  4.9   &  0.0705\\
   &  \hspace{0.30in}$5/2_{2},3/2$   &  2993   &  3061.7   &  13.6818   &  6   &  4.339   &  4.93   &  0.1784\\
   &  \hspace{0.30in}$5/2_{3},3/2$   &  3095   &  3184.54   &  13.1479   &  28   &  4.312   &  4.21   &  0.1898\\
   &  \hspace{0.30in}$1/2_{2},3/2$   &  3360   &  3433   &  4.1747   &  3   &  4.691   &  5.06   &  0.0793\\
   &  \hspace{0.30in}$3/2_{3},3/2$   &  3681   &  3671.7   &  2.5205   &  0.9   &  4.756   &  5.5   &  0.0683\\
   &  \hspace{0.30in}$5/2_{3},3/2$   &  3901   &  3641.5   &  0.3517   &  0.35   &  5.498   &  5.9   &  0.0124\\
   &  \hspace{0.30in}$3/2_{3},3/2$   &  4060   &  3935.2   &  4614   &  0.3   &  4.794   &  5.8   &  0.0625\\
   &  \hspace{0.30in}$5/2_{4},3/2$   &  4166   &  4219.6   &  0.0016   &  0.2   &  7.706   &  5.8   &  0.0001\\
   &  \hspace{0.30in}$1/2_{3},3/2$   &  4326   &  4057   &  0.1354   &  $\leq$ 1   &  5.677   &  > 5.2   &  0.0082\\
   &  \hspace{0.30in}$5/2_{5},3/2$   &  4445   &  4219.6   &  0.0026   &  $\leq$0.20   &  7.328   &  >5.8   &  0.0002\\
   &  \hspace{0.30in}$3/2_{4},3/2$   &   *  &   *  &  0.0001   &   *  &  8.557   &   *  &  0\\
   &  \hspace{0.30in}$1/2_{4},3/2$   &   *  &   *  &  0.0741   &   *  &  5.617   &   *  &  0.0094\\
   &  \hspace{0.30in}$1/2_{5},3/2$   &   *  &   *  &  0.0114   &   *  &  6.085   &   *  &  0.0032\\
   &  \hspace{0.30in}$3/2_{5},3/2$   &   *  &   *  &  0.1018   &   *  &  5.033   &   *  &  0.036\\
   &  \hspace{0.30in}$1/2_{6},3/2$   &   *  &   *  &  0.0013   &   *  &  6.167   &   *  &  0.0026\\
$^{30}$Mg,0,3\cite{30Mg}   &  $^{30}$Al,$1_{1},2$   &  565   &  688   &  85.6711   &  68   &  3.63   &  3.96   &  0.9115\\
   &  \hspace{0.30in}$1_{2},2$   &  2090   &  2413.5   &  14.3289   &  7   &  3.912   &  4.3   &  0.4764\\
$^{31}$Mg,1/2,7/2\cite{31Mg}   &  $^{31}$Al,$1/2_{1},5/2$   &  955   &  946.7   &  29.8291   &  5.2   &  4.337   &  6.1   &  0.1792\\
   &  \hspace{0.30in}$3/2_{1},5/2$   &  1844   &  1613   &  9.5027   &  10.6   &  4.672   &  5.7   &  0.0828\\
   &  \hspace{0.30in}$1/2_{2},5/2$   &  3436   &  3239.3   &  19.7349   &  28.8   &  4.031   &  4.88   &  0.3625\\
   &  \hspace{0.30in}$3/2_{2},5/2$   &  3738   &  3433.3   &  16.0412   &  11.3   &  4.054   &  5.2   &  0.3439\\
   &  \hspace{0.30in}$3/2_{3},5/2$   &  4047   &  3623   &  12.988   &  11.3   &  4.074   &  5.2   &  0.3282\\
   &  \hspace{0.30in}$1/2_{3},5/2$   &  4910   &  4143.3   &  0.0597   &  12.5   &  6.199   &  5   &  0.0025\\
   &  \hspace{0.30in}$3/2_{4},5/2$   &  4939   &  4563.7   &  1.1195   &  7.9   &  4.919   &  5.1   &  0.0469\\
   &  \hspace{0.30in}$3/2_{5},5/2$   &  5293   &  4640.4   &  8.1287   &  1.6   &  3.964   &  5.8   &  0.423\\
   &  \hspace{0.30in}$3/2_{6},5/2$   &  5874   &  4809   &  0.1964   &  1   &  5.417   &  5.9   &  0.0149\\
   &  \hspace{0.30in}$1/2_{4},5/2$   &  5899   &  5046.5   &  0.5841   &  3.5   &  4.936   &  5.3   &  0.0451\\
   &  \hspace{0.30in}$3/2_{7},5/2$   &  6183   &  5149.7   &  0.1885   &  2.9   &  5.342   &  5.4   &  0.0177\\
   &  \hspace{0.30in}$3/2_{8},5/2$   &   *  &   *  &  0.6242   &   *  &  4.712   &   *  &  0.0756\\
   &  \hspace{0.30in}$1/2_{5},5/2$   &   *  &   *  &  0.3546   &   *  &  4.828   &   *  &  0.0578\\
   &  \hspace{0.30in}$3/2_{9},5/2$   &   *  &   *  &  0.0003   &   *  &  7.807   &   *  &  0.0001\\
   &  \hspace{0.30in}$3/2_{10},5/2$   &   *  &   *  &  0.0003   &   *  &  7.606   &   *  &  0.0001\\
   &  \hspace{0.30in}$1/2_{6},5/2$   &   *  &   *  &  0.0191   &   *  &  5.753   &   *  &  0.0069\\
   &  \hspace{0.30in}$1/2_{7},5/2$   &   *  &   *  &  0.6088   &   *  &  4.153   &   *  &  0.2737\\
   &  \hspace{0.30in}$1/2_{8},5/2$   &   *  &   *  &  0.0052   &   *  &  5.822   &   *  &  0.0059\\
   &  \hspace{0.30in}$1/2_{9},5/2$   &   *  &   *  &  0.0128   &   *  &  5.138   &   *  &  0.0283\\
   &  \hspace{0.30in}$1/2_{10},5/2$   &   *  &   *  &  0.0019   &   *  &  5.846   &   *  &  0.0055\\
$^{32}$Mg,0,4\cite{32Mg}   &  $^{32}$Al,$1_{1},3$   &  0   &  0   &  99.2926   &  55   &  3.385   &  4.4   &  1.604\\
   &  \hspace{0.30in}$1_{2},3$   &  2997   &  2765.3   &  0.7073   &  24.6   &  4.826   &  4.1   &  0.05\\
   &  \hspace{0.30in}$1_{3},3$   &  5466   &  3202.2   &  0.0001   &  10.7   &  7.824   &  4.4   &  0.0001\\
$^{28}$Al,3,1\cite{nndc}   &  $^{28}$Si,$2_{1},0$   &  1932   &  1778.987   &  99.6377   &  99.99   &  4.605   &  4.8664   &  0.0803\\
   &  \hspace{0.30in}$4_{1},0$   &   *  &   *  &  0.3605   &   *  &  5.488   &   *  &  0.0126\\
   &  \hspace{0.30in}$3_{1},0$   &   *  &   *  &  0.0018   &   *  &  4.359   &   *  &  0.1701\\
$^{29}$Al,5/2,3/2\cite{29Al}   &  $^{29}$Si,$3/2_{1},1/2$   &  1285   &  1273.391   &  90.9415   &  89.9   &  4.638   &  5.05   &  0.0896\\
   &  \hspace{0.30in}$5/2_{1},1/2$   &  2063   &  2028.17   &  2.7908   &  3.8   &  5.695   &  5.733   &  0.0078\\
   &  \hspace{0.30in}$3/2_{2},1/2$   &  2525   &  2425.99   &  6.1117   &  6.3   &  5.029   &  5.026   &  0.0364\\
   &  \hspace{0.30in}$5/2_{2},1/2$   &  3356   &  3067.28   &  0.1558   &  0.033   &  5.863   &  6.11   &  0.0053\\
   &  \hspace{0.30in}$7/2_{1},1/2$   &   *  &   *  &  0.0002   &   *  &  7.492   &   *  &  0.0001\\
   &  \hspace{0.30in}$7/2_{2},1/2$   &   *  &   *  &  0   &   *  &  4.751   &   *  &  0.069\\
$^{30}$Al,3,2\cite{30Al}   &  $^{30}$Si,$2_{1},1$   &  2266   &  2235.326   &  8.6243   &  17.1   &  5.52   &  5.619   &  0.0118\\
   &  \hspace{0.30in}$2_{2},1$   &  3506   &  3498.5   &  63.6615   &  67.3   &  4.333   &  4.578   &  0.1808\\
   &  \hspace{0.30in}$3_{1},1$   &  4869   &  4810.31   &  12.2118   &  5.7   &  4.635   &  5.06   &  0.0901\\
   &  \hspace{0.30in}$2_{3},1$   &  4866   &  4830.84   &  7.0753   &  6.6   &  4.871   &  4.985   &  0.0523\\
   &  \hspace{0.30in}$3_{2},1$   &  5131   &  5231.56   &  4.2795   &  2.6   &  5   &  5.17   &  0.0389\\
   &  \hspace{0.30in}$4_{1},1$   &  5334   &  5951.6   &  0.2355   &  0.16   &  6.186   &  5.92   &  0.0025\\
   &  \hspace{0.30in}$4_{2},1$   &   *  &   *  &  0.6173   &   *  &  5.564   &   *  &  0.0106\\
   &  \hspace{0.30in}$2_{4},1$   &  5919   &  5614.02   &  1.1   &  0.3   &  5.293   &  5.87   &  0.0198\\

                                      \hline
    \end{tabular}
  \end{center}
\end{table*}

\addtocounter{table}{-1}

\begin{table*}
  \begin{center}
    \leavevmode
    \caption{{\em Continuation.\/}}   

    \label{tab:mgt_comp} 
   \begin{tabular}{ccccccccc}
        \hline \hline
            & &  \multicolumn{2}{c}{Energy level }  &\multicolumn{2}{c}{Branching Ratio } & \multicolumn{2}{c}{$log ft$ } & B(GT)\\
        \cline{3-4} \cline{7-8}
        Parent (J$_{i}$,T$_{i})[Ref.]$ & Daughter J$_{f}$,T$_{f}$& Th. & Expt. &  Th.(\%) &  Expt.(\%) &  (Th). & (Exp.).&   (Th.)\\
         \hline   &  \hspace{0.30in}$3_{3},1$   &   *  &   *  &  0.2221   &   *  &  5.523   &   *  &  0.0117\\
   &  \hspace{0.30in}$4_{3},1$   &   *  &   *  &  1.2051   &   *  &  4.724   &   *  &  0.0735\\
   &  \hspace{0.30in}$3_{4},1$   &   *  &   *  &  0.0594   &   *  &  5.939   &   *  &  0.0045\\
   &  \hspace{0.30in}$4_{4},1$   &   *  &   *  &  0.7082   &   *  &  4.401   &   *  &  0.1544\\
$^{31}$Al,5/2,5/2\cite{31Al}   &  $^{31}$Si,$3/2_{1},3/2$   &  0   &  0   &  63.5626   &   *  &  4.465   &   *  &  0.1335\\
   &  \hspace{0.30in}$5/2_{1},3/2$   &  1574   &  1694.83   &  12.5331   &   *  &  4.725   &   *  &  0.0732\\
   &  \hspace{0.30in}$3/2_{2},3/2$   &  2252   &  2316.7   &  23.6703   &   *  &  4.225   &   *  &  0.2317\\
   &  \hspace{0.30in}$5/2_{2},3/2$   &  2805   &  2787.7   &  0.1089   &   *  &  6.36   &   *  &  0.0017\\
   &  \hspace{0.30in}$7/2_{1},3/2$   &   *  &   *  &  0.0974   &   *  &  6.014   &   *  &  0.0038\\
   &  \hspace{0.30in}$7/2_{2},3/2$   &   *  &   *  &  0.0277   &   *  &  6.075   &   *  &  0.0033\\
$^{32}$Al,1,3\cite{32Al}   &  $^{32}$Si,$0_{1},2$   &  0   &  0   &  83.1797   &  85   &  4.191   &   *  &  0.2507\\
   &  \hspace{0.30in}$2_{1},2$   &  2053   &  1941.4   &  1.0148   &  4.7   &  5.748   &   *  &  0.0069\\
   &  \hspace{0.30in}$2_{2},2$   &  4239   &  4230.8   &  3.7196   &  3   &  4.725   &   *  &  0.0732\\
   &  \hspace{0.30in}$0_{2},2$   &  4990   &  4983.9   &  10.5612   &  4.3   &  4.089   &   *  &  0.3171\\
   &  \hspace{0.30in}$1_{1},2$   &  5427   &  5785.7   &  0.2391   &  1.7   &  5.62   &   *  &  0.0093\\
   &  \hspace{0.30in}$1_{2},2$   &   *  &   *  &  1.2855   &   *  &  4.219   &   *  &  0.2349\\
$^{33}$Al,5/2,7/2\cite{33Al}   &  $^{33}$Si,$3/2_{1},5/2$   &  0   &  0   &  94.8585   &  88   &  3.874   &  4.3   &  0.5199\\
   &  \hspace{0.30in}$7/2_{1},5/2$   &   *  &   *  &  0.2942   &   *  &  5.504   &   *  &  0.0122\\
   &  \hspace{0.30in}$5/2_{1},5/2$   &  4595   &  4341   &  0.8076   &  1.3   &  4.958   &  5.2   &  0.0429\\
   &  \hspace{0.30in}$3/2_{2},5/2$   &   *  &   *  &  3.2313   &   *  &  4.304   &   *  &  0.1934\\
   &  \hspace{0.30in}$5/2_2{},5/2$   &   *  &   *  &  0.1936   &   *  &  5.119   &   *  &  0.0296\\
   &  \hspace{0.30in}$7/2_{2},5/2$   &   *  &   *  &  0.6147   &   *  &  4.303   &   *  &  0.1938\\
$^{31}$Si,3/2,3/2\cite{31Si}   &  $^{31}$P,$1/2_{1},1/2$   &  0   &  0   &  99.4932   &  99.9446   &  5.312   &  5.525   &  0.019\\
   &  \hspace{0.30in}$3/2_{1},1/2$   &  1173   &  1266.1   &  0.5068   &  0.0554   &  5.086   &  5.747   &  0.0319\\
   &  \hspace{0.30in}$5/2_{1},1/2$   &   *  &   *  &  0   &   *  &  4.079   &   *  &  0.3244\\
$^{32}$Si,1,2\cite{nndc}   &  $^{32}$P,$1_{1},1$   &  0   &  0   &  92.3402   &  100   &  4.494   &  8.21   &  0.1247\\
   &  \hspace{0.30in}$2_{1},1$   &   *  &   *  &  7.6598   &   *  &  4.499   &   *  &  0.1232\\
   &  \hspace{0.30in}$0_{1},1$   &   *  &   *  &  0   &   *  &  5.378   &   *  &  0.0163\\
   &  \hspace{0.30in}$1_{2},1$   &   *  &   *  &  0   &   *  &  5.193   &   *  &  0.0249\\
   &  \hspace{0.30in}$2_{2},1$   &   *  &   *  &  0   &   *  &  7.394   &   *  &  0.0002\\
   &  \hspace{0.30in}$0_{2},1$   &   *  &   *  &  0   &   *  &  5.812   &   *  &  0.006\\
$^{32}$P,0,1\cite{32P}   &  $^{32}$P,$1_{1},0$   &  0   &  0   &  100   &  100   &  4.256   &  7.9002   &  0.216\\
$^{33}$P,3/2,3/2\cite{33P}   &  $^{33}$P,$3/2_{1},1/2$   &  0   &  0   &  100   &  100   &  5.35   &  5.022   &  0.0174\\
   &  \hspace{0.30in}$1/2_{1},1/2$   &   *  &   *  &  0   &   *  &  5.397   &   *  &  0.0156\\
   &  \hspace{0.30in}$5/2_{1},1/2$   &   *  &   *  &  0   &   *  &  4.653   &   *  &  0.0864\\
$^{34}$P,1,2\cite{34P}   &  $^{34}$P,$1_{1},1$   &   *  &   *  &  60.2666   &  84.8   &  4.743   &   *  &  0.0704\\
   &  \hspace{0.30in}$1_{2},1$   &   *  &   *  &  39.7334   &   *  &  4.203   &   *  &  0.2436\\
$^{35}$P,1/2,5/2\cite{35P}   &  $^{35}$P,$3/2_{1},3/2$   &  0   &  0   &  7.7626   &   *  &  6.012   &   *  &  0.0038\\
   &  \hspace{0.30in}$1/2_{1},3/2$   &  1676   &  1572.29   &  90.7258   &  98.8   &  3.906   &  4.122   &  0.4834\\
   &  \hspace{0.30in}$3/2_{2},3/2$   &  2721   &  2704   &  1.5108   &   *  &  4.488   &   *  &  0.1264\\
   &  \hspace{0.30in}$3/2_{3},3/2$   &  2802   &  2938.4   &  0   &   *  &  4.275   &   *  &  0.2065\\
   &  \hspace{0.30in}$3/2_{4},3/2$   &  3374   &   *  &  0   &   *  &  4.555   &   *  &  0.1085\\
   &  \hspace{0.30in}$1/2_{2},3/2$   &   *  &   *  &  0   &   *  &  4.052   &   *  &  0.3451\\
   &  \hspace{0.30in}$1/2_{3},3/2$   &   *  &   *  &  0   &   *  &  6.326   &   *  &  0.0018\\
   &  \hspace{0.30in}$1/2_{4},3/2$   &   *  &   *  &  0   &   *  &  3.839   &   *  &  0.5638\\
$^{36}$Cl(0,1)\cite{nndc}   &  $^{36}$Ar,$1_{1},0$   &  0   &  0   &  99.9742   &  98.1   &  5.16   &  13.321   &  0.0269\\
   &  \hspace{0.30in}$1_{2},0$   &   *  &   *  &  0.667   &   *  &  4.682   &   *  &  0.0808\\

                                      \hline
    \end{tabular}
  \end{center}
\end{table*}

\begin{figure*}
\begin{center}
\resizebox{1.0\textwidth}{!}{
\includegraphics{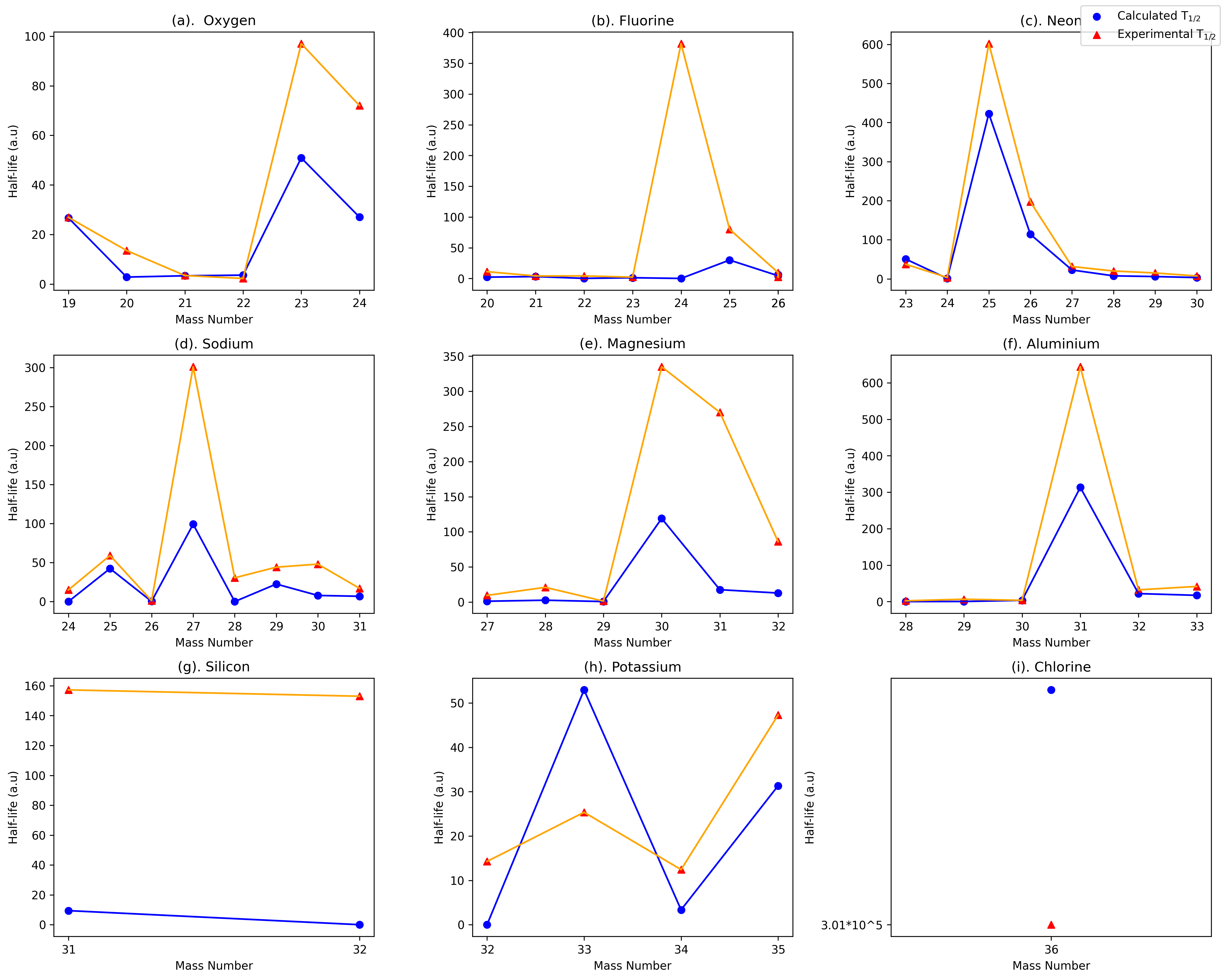} }
\end{center}
\caption{
Comparison of experimental and theoretical half-life values with theoretical calculations. Each triangle represents an experimental half-life, while each circle represents a theoretical half-life. }
\label{Table1}
\end{figure*}

\begin{table*}
 \begin{center}
    \leavevmode
    \caption{Theoretical $B(GT)$, $M(GT)$  matrix elements. The calculated shell model results with USDB interaction.
    We have taken experimental data from ref. \cite{nndc}. Here
  $I_\beta^- $ are the branching ratios.}
    \label{tab:mgt_comp} 
    \begin{tabular}{lcccccrrrr}
    \hline\hline
Parent &    Daughter&  $J_i, T_i$ & $J_f, T_f$   &    I$_{\beta^-}$&    B(GT) &    Sum B(GT)&    M(GT)&    W&    R(GT)\\ \hline
$^{19}$O&    $^{19}$F&    2.5,1.5&    5/2,1/2&    20.9872&    0.0062&    0.0062&    0.193 &    8.262&    0.023\\
&    &    &    3/2,1/2&    79.0128&    0.1627&    0.1689&    0.988 &    &    0.120\\
$^{20}$O&    $^{20}$F&    0,2&    1,1&    97.7005&    0.7674&    0.7674&    0.876 &    3.895&    0.225\\
&    &    &    1,1&    2.2995&    1.035&    1.8024&    1.017 &    &    0.261\\
$^{21}$O&    $^{21}$F&    2.5,2.5&    5/2,3/2&    8.0681&    0.0017&    0.0017&    0.101 &    10.666&    0.009\\
&    &    &    3/2,3/2&    33.1716&    0.0226&    0.0243&    0.368 &    &    0.035\\
&    &    &    5/2,3/2&    29.6415&    0.0854&    0.1097&    0.716 &    &    0.067\\
&    &    &    3/2,3/2&    15.4338&    0.0473&    0.1571&    0.533 &    &    0.050\\
&    &    &    7/2,3/2&    1.5963&    0.005&    0.01621&    0.173 &    &    0.016\\
&    &    &    5/2,3/2&    0.8667&    0.0053&    0.1674&    0.178 &    &    0.017\\
&    &    &    3/2,3/2&    5.0825&    0.0397&    0.207&    0.488 &    &    0.046\\
&    &    &    7/2,3/2&    6.1384&    0.0508&    0.2579&    0.552 &    &    0.052\\
&    &    &    7/2,3/2&    0.0012&    0&    0.2579&    0.000 &    &    0.000\\
$^{2}$O&    $^{22}$F&    0,3&    1,2&    18.5581&    0.1027&    0.1027&    0.320 &    4.770&    0.067\\
&    &    &    1,2&    81.4419&    0.961&    1.0637&    0.980 &    &    0.206\\
$^{23}$O&    $^{23}$F&    0.5,3.5&    1/2,5/2&    43.7428&    0.1239&    0.1239&    0.498 &    7.286&    0.068\\
&    &    &    3/2,5/2&    25.2553&    0.1792&    0.3031&    0.599 &    &    0.082\\
&    &    &    3/2,5/2&    3.5176&    0.0325&    0.3356&    0.255 &    &    0.035\\
&    &    &    1/2,5/2&    0.0322&    0.0005&    0.3361&    0.032 &    &    0.004\\
&    &    &    3/2,5/2&    5.7789&    0.1423&    0.4784&    0.533 &    &    0.073\\
&    &    &    1/2,5/2&    8.7683&    0.275&    0.7534&    0.742 &    &    0.102\\
&    &    &    3/2,5/2&    9.2429&    0.3036&    1.057&    0.779 &    &    0.107\\
&    &    &    3/2,5/2&    0.8655&    0.036&    1.093&    0.268 &    &    0.037\\
&    &    &    1/2,5/2&    2.7964&    0.324&    1.417&    0.805 &    &    0.110\\
$^{24}$O&    $^{24}$F&    0,4&    1,3&    100&    0.6174&    0.6174&    0.786 &    5.152&    0.153\\
$^{20}$F&    $^{20}$Ne&    2,1&    2,0&    99.7575&    0.0594&    0.0594&    0.545 &    6.158&    0.089\\
&    &    &    2,0&    0.2425&    0.3028&    0.3622&    1.230 &    &    0.200\\
$^{21}$F&    $^{21}$Ne&    2.5,1.5&    3/2,1/2&    12.8035&    0.0145&    0.0145&    0.295 &    8.262&    0.036\\
&    &    &    5/2,1/2&    71.3677&    0.1011&    0.1156&    0.779 &    &    0.094\\
&    &    &    7/2,1/2&    15.7346&    0.0967&    0.2123&    0.762 &    &    0.092\\
&    &    &    5/2,1/2&    0.0059&    0.0007&    0.2131&    0.065 &    &    0.008\\
&    &    &    5/2,1/2&    0.0552&    0.0894&    0.3024&    0.732 &    &    0.089\\
&    &    &    3/2,1/2&    0.0329&    0.1827&    0.4851&    1.047 &    &    0.127\\
&    &    &    7/2,1/2&    0.0001&    0.0202&    0.5053&    0.348 &    &    0.042\\
$^{22}$F&    $^{22}$Ne&    4,2&    4,1&    1.2189&    0.0008&    0.0008&    0.085 &    11.684&    0.007\\
&    &    &    4,1&    52.8062&    0.0965&    0.0973&    0.932 &    &    0.080\\
&    &    &    3,1&    16.7565&    0.0322&    0.1295&    0.538 &    &    0.046\\
&    &    &    4,1&    3.7067&    0.0115&    0.141&    0.322 &    &    0.028\\
&    &    &    3,1&    0.7471&    0.0026&    0.1437&    0.153 &    &    0.013\\
&    &    &    5,1&    18.7559&    0.1093&    0.253&    0.992 &    &    0.085\\
&    &    &    3,1&    4.8036&    0.004&    0.2569&    0.190 &    &    0.016\\
&    &    &    5,1&    0.7451&    0.0742&    0.3312&    0.817 &    &    0.070\\
$^{23}$F&    $^{23}$Ne&    2.5,2.5&    5/2,3/2&    34.1604&    0.0149&    0.0149&    0.299 &    10.666&    0.028\\
&    &    &    7/2,3/2&    0.6805&    0.0158&    0.0158&    0.308 &    &    0.029\\
&    &    &    3/2,3/2&    5.0902&    0.023&    0.023&    0.371 &    &    0.035\\
&    &    &    5/2,3/2&    1.2735&    0.0255&    0.0255&    0.391 &    &    0.037\\
&    &    &    3/2,3/2&    17.6043&    0.104&    0.104&    0.790 &    &    0.074\\
&    &    &    5/2,3/2&    27.8797&    0.283&    0.283&    1.303 &    &    0.122\\
&    &    &    7/2,3/2&    1.5773&    0.2937&    0.2937&    1.327 &    &    0.124\\
&    &    &    3/2,3/2&    0.5801&    0.2979&    0.2979&    1.337 &    &    0.125\\
&    &    &    7/2,3/2&    10.4902&    0.4276&    0.4276&    1.602 &    &    0.150\\
$^{24}$F&    $^{24}$Ne&    3,3&    2,2&    22.8166&    0.0111&    0.0111&    0.279 &    12.620&    0.022\\
&    &    &    2,2&    21.922&    0.0198&    0.0309&    0.372 &    &    0.030\\
&    &    &    4,2&    44.8041&    0.0446&    0.0755&    0.559 &    &    0.044\\
&    &    &    3,2&    4.4049&    0.0062&    0.0817&    0.208 &    &    0.017\\
&    &    &    3,2&    0.2183&    0.0004&    0.0821&    0.053 &    &    0.004\\
&    &    &    4,2&    5.8341&    0.0123&    0.0944&    0.293 &    &    0.023\\
$^{25}$F&    $^{25}$Ne&    2.5,3.5&    3/2,5/2&    19.6223&    0.0404&    0.0404&    0.492 &    12.620&    0.039\\
&    &    &    3/2,5/2&    3.2086&    0.01&    0.0505&    0.245 &    &    0.019\\
&    &    &    7/2,5/2&    30.4252&    0.1211&    0.1716&    0.852 &    &    0.068\\
&    &    &    7/2,5/2&    30.4252&    0.1211&    0.2927&    0.852 &    &    0.068\\
&    &    &    7/2,5/2&    0.327&    0.0017&    0.2944&    0.101 &    &    0.008\\
&    &    &    7/2,5/2&    0.327&    0.0017&    0.2961&    0.101 &    &    0.008\\
&    &    &    3/2,5/2&    2.3347&    0.0134&    0.3095&    0.284 &    &    0.022\\                          
\hline
 \end{tabular}
   \end{center}
     \end{table*}

\addtocounter{table}{-1}

\begin{table*}
  \begin{center}
    \leavevmode
    \caption{{\em Continuation.\/}}   

    \label{tab:mgt_comp} 
     \begin{tabular}{lcccccrrrr}
    \hline\hline
Parent &    Daughter&  $J_i, T_i$ & $J_f, T_f$   &    I$_{\beta^-}$&    B(GT) &    Sum B(GT)&    M(GT)&    W&    R(GT)\\ \hline
$^{26}$F &    $^{26}$Ne&    4,4&    4,3&    73.6897&    0.1609&    0.1609&    1.203 & 16.497
   &  0.073 \\ 
&    &    &    3,3&    3.3661&    0.0123&    0.1732&    0.111 &    & 0.007   \\
&    &    &    4,3&    13.3773&    0.0505&    0.2237&    0.225 &    & 0.014   \\
&    &    &    3,3&    1.9842&    0.002&    0.2329&    0.045 &    &  0.003  \\
&    &    &    5,3&    1.7723&    0.0145&    0.2474&    0.120 &    & 0.007   \\
&    &    &    5,3&    5.8103&    0.0546&    0.302&    0.234 &    &  0.014  \\
$^{26}$F &    $^{26}$Ne&    1,4&    0,3&    35.2724&    0.1129&    0.1129&    0.582 & 9.5247  &  0.061 \\ 
&    &    &    2,3&    48.2307&    0.2755&    0.3884&    0.525 &    &  0.055  \\
&    &    &    2,3&    10.7805&    0.1053&    0.4937&    0.324 &    &  0.034  \\
&    &    &    0,3&    4.2856&    0.0564&    0.5501&    0.237 &    &  0.025  \\
&    &    &    1,3&    1.2559&    0.0348&    0.5873&    0.187 &    &  0.020  \\
$^{23}$Ne&    $^{23}$Na&    2.5,1.5&    3/2,1/2&    57.9673&    0.0132&    0.0132&    0.281 &    16.523&    0.017\\
&    &    &    5/2,1/2&    41.2246&    0.0145&    0.0277&    0.295 &    &    0.018\\
&    &    &    7/2,1/2&    0.3014&    0.0014&    0.0291&    0.092 &    &    0.006\\
&    &    &    3/2,1/2&    0.5064&    0.008&    0.0371&    0.219 &    &    0.013\\
$^{24}$Ne&    $^{24}$Na&    0,2&    1,1&    84.8747&    0.2052&    0.2052&    0.453 &    3.895&    0.116\\
&    &    &    1,1&    15.1253&    0.1851&    0.3903&    0.430 &    &    0.110\\
$^{25}$Ne &    $^{25}$Na&    0.5,2.5&    3/2,3/2&    69.0255&    0.1776&    0.1776&    0.596 &    6.158&    0.097\\
&    &    &    1/2,3/2&    24.4595&    0.1123&    0.2899&    0.474 &    &    0.077\\
&    &    &    3/2,3/2&    4.1392&    0.042&    0.3319&    0.290 &    &    0.047\\
&    &    &    3/2,3/2&    1.1953&    0.0556&    0.3875&    0.333 &    &    0.054\\
&    &    &    1/2,3/2&    1.1402&    0.1116&    0.4991&    0.472 &    &    0.077\\
$^{26}$Ne &    $^{26}$Na&    0,3&    1,1&    83.4445&    0.776&    0.776&    0.881 &    4.770&    0.185\\
&    &    &    1,1&    12.4313&    0.2818&    1.0578&    0.531 &    &    0.111\\
&    &    &    1,1&    3.9369&    0.2393&    1.2971&    0.489 &    &    0.103\\
$^{27}$Ne&    $^{27}$Na&    1.5,3.5&    5/2,5/2&    64.4845&    0.2381&    0.2381&    0.976 &    10.304&    0.095\\
&    &    &    3/2,5/2&    14.9302&    0.0557&    0.2938&    0.472 &    &    0.046\\
&    &    &    1/2,5/2&    0.4334&    0.032&    0.297&    0.358 &    &    0.035\\
&    &    &    5/2,5/2&    6.9833&    0.0921&    0.389&    0.607 &    &    0.059\\
&    &    &    3/2,5/2&    10.2101&    0.1545&    0.5435&    0.786 &    &    0.076\\
&    &    &    1/2,5/2&    1.4067&    0.0484&    0.625&    0.440 &    &    0.043\\
$^{28}$Ne&    $^{28}$Na&    0,4&    1,3&    72.7406&    0.8842&    0.8842&    0.940 &    5.508&    0.171\\
&    &    &    1,3&    14.7253&    0.4199&    1.3041&    0.648 &    &    0.118\\
&    &    &    1,3&    1.9715&    0.0733&    1.3774&    0.271 &    &    0.049\\
&    &    &    1,3&    2.9957&    0.14&    1.5174&    0.374 &    &    0.068\\
&    &    &    1,3&    7.4902&    0.6772&    2.1946&    0.823 &    &    0.149\\
&    &    &    1,3&    0.0766&    0.0075&    2.2021&    0.087 &    &    0.016\\
$^{29}$Ne &    $^{29}$Na&    1.5,4.5&    5/2,7/2&    24.0716&    0.1173&    0.1173&    0.685 &    11.684&    0.059\\
&    &    &    5/2,7/2&    24.0716&    0.1173&    0.2346&    0.685 &    &    0.059\\
&    &    &    1/2,7/2&    1.5023&    0.0148&    0.2494&    0.243 &    &    0.021\\
&    &    &    5/2,7/2&    21.6126&    0.3329&    0.5823&    1.154 &    &    0.099\\
&    &    &    5/2,7/2&    21.6126&    0.3329&    0.9152&    1.154 &    &    0.099\\
&    &    &    1/2,7/2&    3394&    0.0055&    0.9207&    0.148 &    &    0.013\\
&    &    &    1/2,7/2&    5.0929&    0.1004&    1.0211&    0.634 &    &    0.054\\
$^{31}$Ne &    $^{30}$Na&    0,5&    1,4&    60.3304&    0.7586&    0.7586&    0.871 &    6.158&    0.141\\
&    &    &    1,4&    23.1876&    0.7233&    1.4819&    0.850 &    &    0.138\\
&    &    &    1,4&    16.482&    0.789&    2.2709&    0.888 &    &    0.144\\
$^{24}$Na&    $^{24}$Mg&    4,0&    4,0&    76.2805&    0.0079&    0.0079&    0.267 &    8.262&    0.032\\
&    &    &    3,0&    19.7423&    0.0038&    0.0117&    0.185 &    &    0.022\\
&    &    &    4,0&    3.9619&    0.0018&    0.0134&    0.127 &    &    0.015\\
$^{25}$Na&    $^{25}$Mg&    2.5,1.5&    5/2,1/2&    68.0102&    0.0327&    0.0327&    0.443 &    8.262&    0.054\\
&    &    &    3/2,1/2&    24.5795&    0.0526&    0.0853&    0.562 &    &    0.068\\
&    &    &    7/2,1/2&    6.6518&    0.0433&    0.1287&    0.510 &    &    0.062\\
&    &    &    5/2,1/2&    0.3719&    0.0044&    0.133&    0.162 &    &    0.020\\
&    &    &    3/2,1/2&    0.385&    0.049&    0.1821&    0.542 &    &    0.066\\
$^{26}$Na&    $^{26}$Mg&    3,2&    2,1&    76.1699&    0.1036&    0.1036&    0.852 &    10.304&    0.083\\
&    &    &    2,1&    0.5895&    0.0015&    0.1051&    0.102 &    &    0.010\\
&    &    &    3,1&    1.108&    0.0046&    0.1097&    0.179 &    &    0.017\\
&    &    &    3,1&    8.2572&    0.0459&    0.1555&    0.567 &    &    0.055\\
&    &    &    4,1&    0.8785&    0.005&    0.1606&    0.187 &    &    0.018\\
&    &    &    2,1&    2.5027&    0.0152&    0.1758&    0.326 &    &    0.032\\
&    &    &    2,1&    3.3914&    0.028&    0.2038&    0.443 &    &    0.043\\
&    &    &    4,1&    0.0487&    0.0004&    0.2042&    0.053 &    &    0.005\\
&    &    &    2,1&    0.0186&    0.0002&    0.2044&    0.037 &    &    0.004\\
&    &    &    4,1&    0.2617&    0.0035&    0.2079&    0.157 &    &    0.015\\
\hline
    \end{tabular}
  \end{center}
\end{table*}

\addtocounter{table}{-1}

\begin{table*}
  \begin{center}
    \leavevmode
    \caption{{\em Continuation.\/}}   

    \label{tab:mgt_comp} 
     \begin{tabular}{lcccccrrrr}
    \hline\hline
Parent &    Daughter&  $J_i, T_i$ & $J_f, T_f$   &    I$_{\beta^-}$&    B(GT) &    Sum B(GT)&    M(GT)&    W&    R(GT)\\ \hline
$^{27}$Na&    $^{27}$Mg&    2.5,2.5&    3/2,3/2&    83.1719&    0.3204&    0.3204&    1.387 &    10.666&    0.130\\
&    &    &    5/2,3/2&    6.8371&    0.038&    0.3584&    0.477 &    &    0.045\\
&    &    &    5/2,3/2&    1.3168&    0.0083&    0.3668&    0.223 &    &    0.021\\
&    &    &    7/2,3/2&    0.6022&    0.0079&    0.3747&    0.218 &    &    0.020\\
&    &    &    3/2,3/2&    0.3861&    0.0066&    0.3813&    0.199 &    &    0.019\\
&    &    &    7/2,3/2&    0.956&    0.0164&    0.3977&    0.314 &    &    0.029\\
&    &    &    3/2,3/2&    0.8395&    0.0159&    0.4135&    0.309 &    &    0.029\\
&    &    &    5/2,3/2&    0.3604&    0.01&    0.4235&    0.245 &    &    0.023\\
&    &    &    5/2,3/2&    0.7471&    0.0253&    0.4488&    0.390 &    &    0.037\\
&    &    &    7/2,3/2&    0.5299&    0.024&    0.4728&    0.379 &    &    0.036\\
&    &    &    5/2,3/2&    1.6929&    0.0933&    0.5661&    0.748 &    &    0.070\\
$^{28}$Na&    $^{28}$Mg&    1,3&    0,2&    59.6477&    0.1483&    0.1483&    0.667 &    8.262&    0.081\\
&    &    &    2,2&    4.2164&    0.0181&    0.1664&    0.233 &    &    0.028\\
&    &    &    0,2&    21.1598&    0.2616&    0.428&    0.886 &    &    0.107\\
&    &    &    2,2&    2.3639&    0.0379&    0.4659&    0.337 &    &    0.041\\
&    &    &    1,2&    3.5429&    0.0604&    0.5263&    0.426 &    &    0.052\\
&    &    &    2,2&    0.0136&    0.0002&    0.5265&    0.024 &    &    0.003\\
&    &    &    1,2&    2.1012&    0.0563&    0.5828&    0.411 &    &    0.050\\
&    &    &    2,2&    0.1946&    0.054&    0.5882&    0.402 &    &    0.049\\
&    &    &    2,2&    1.3337&    0.0489&    0.6371&    0.383 &    &    0.046\\
&    &    &    0,2&    0.3256&    0.0164&    0.6535&    0.222 &    &    0.027\\
&    &    &    2,2&    0.2251&    0.0143&    0.6678&    0.207 &    &    0.025\\
&    &    &    1,2&    0.6703&    0.0457&    0.7135&    0.370 &    &    0.045\\
&    &    &    0,2&    0.4852&    0.0347&    0.7482&    0.323 &    &    0.039\\
&    &    &    1,2&    0.139&    0.0126&    0.7608&    0.194 &    &    0.024\\
&    &    &    2,2&    0.1039&    0.0103&    0.7711&    0.176 &    &    0.021\\
&    &    &    2,2&    2.7145&    0.2843&    1.0554&    0.924 &    &    0.112\\
$^{29}$Na&    $^{29}$Mg&    1.5,3.5&    3/2,5/2&    35.1635&    0.0974&    0.0974&    0.624 &    10.304&    0.061\\
&    &    &    1/2,5/2&    11.581&    0.0326&    0.13&    0.361 &    &    0.035\\
&    &    &    5/2,5/2&    0.1664&    0.008&    0.1308&    0.179 &    &    0.017\\
&    &    &    3/2,5/2&    0.5719&    0.004&    0.1348&    0.126 &    &    0.012\\
&    &    &    1/2,5/2&    28.3018&    0.2242&    0.359&    0.947 &    &    0.092\\
&    &    &    5/2,5/2&    4.4167&    0.0445&    0.4035&    0.422 &    &    0.041\\
&    &    &    3/2,5/2&    10.6317&    0.1265&    0.53&    0.711 &    &    0.069\\
&    &    &    5/2,5/2&    3.8083&    0.0468&    0.5767&    0.433 &    &    0.042\\
&    &    &    5/2,5/2&    0.3995&    0.007&    0.5837&    0.167 &    &    0.016\\
&    &    &    3/2,5/2&    4.8335&    0.001&    0.5838&    0.063 &    &    0.006\\
&    &    &    1/2,5/2&    0.1232&    0.1623&    0.7461&    0.806 &    &    0.078\\
&    &    &    1/2,5/2&    9.4971&    0.0062&    0.7523&    0.157 &    &    0.015\\
$^{30}$Na&    $^{30}$Mg&    2,4&    2,3&    10.3898&    0.021&    0.021&    0.324 &    12.316&    0.026\\
&    &    &    2,3&    0.6652&    0.0393&    0.0602&    0.443 &    &    0.036\\
&    &    &    3,3&    28.2814&    0.0037&    0.0639&    0.136 &    &    0.011\\
&    &    &    2,3&    1.2435&    0.1656&    0.2295&    0.910 &    &    0.074\\
&    &    &    2,3&    7.4875&    0.0082&    0.2378&    0.202 &    &    0.016\\
&    &    &    1,3&    8.878&    0.0499&    0.2877&    0.499 &    &    0.041\\
&    &    &    1,3&    5.8731&    0.0715&    0.3592&    0.598 &    &    0.049\\
&    &    &    3,3&    6.0893&    0.0657&    0.4249&    0.573 &    &    0.047\\
&    &    &    2,3&    0.0782&    0.0732&    0.4981&    0.605 &    &    0.049\\
&    &    &    2,3&    5.1763&    0.001&    0.499&    0.071 &    &    0.006\\
&    &    &    3,3&    0.6439&    0.0692&    0.5682&    0.588 &    &    0.048\\
&    &    &    2,3&    1.1474&    0.0095&    0.5777&    0.218 &    &    0.018\\
&    &    &    3,3&    3.2149&    0.0179&    0.5957&    0.299 &    &    0.024\\
&    &    &    1,3&    0.4182&    0.0524&    0.6481&    0.512 &    &    0.042\\
&    &    &    2,3&    2.7423&    0.077&    0.6559&    0.620 &    &    0.050\\
&    &    &    2,3&    0.0097&    0.0586&    0.7144&    0.541 &    &    0.044\\
&    &    &    3,3&    2.5133&    0.0002&    0.7146&    0.032 &    &    0.003\\
&    &    &    3,3&    0.1351&    0.0662&    0.7809&    0.575 &    &    0.047\\
&    &    &    1,3&    1.793&    0.0037&    0.7846&    0.136 &    &    0.011\\
&    &    &    3,3&    0.0078&    0.0548&    0.8394&    0.523 &    &    0.043\\
&    &    &    1,3&    0.0809&    0.0002&    0.8396&    0.032 &    &    0.003\\
&    &    &    3,3&    0.0134&    0.0029&    0.8425&    0.120 &    &    0.010\\
&    &    &    1,3&    2.2968&    0.0005&    0.843&    0.050 &    &    0.004\\
\hline
    \end{tabular}
  \end{center}
\end{table*}
\addtocounter{table}{-1}

\begin{table*}
  \begin{center}
    \leavevmode
    \caption{{\em Continuation.\/}}   

    \label{tab:mgt_comp} 
     \begin{tabular}{lcccccrrrr}
    \hline\hline
Parent &    Daughter&  $J_i, T_i$ & $J_f, T_f$   &    I$_{\beta^-}$&    B(GT) &    Sum B(GT)&    M(GT)&    W&    R(GT)\\ \hline

$^{31}$Na&    $^{31}$Mg&    1.5,4.5&    3/2,7/2&    9.69&    0.0439&    0.0439&    0.419 &    11.684&    0.036\\
&    &    &    5/2,7/2&    9.365&    0.0668&    0.1107&    0.517 &    &    0.044\\
&    &    &    1/2,7/2&    9.5712&    0.0969&    0.2076&    0.623 &    &    0.053\\
&    &    &    5/2,7/2&    24.8454&    0.3223&    0.5299&    1.135 &    &    0.097\\
&    &    &    3/2,7/2&    14.5603&    0.2101&    0.74&    0.917 &    &    0.078\\
&    &    &    1/2,7/2&    16.7288&    0.3047&    1.0447&    1.104 &    &    0.094\\
&    &    &    5/2,7/2&    5.7916&    0.1287&    1.1734&    0.717 &    &    0.061\\
&    &    &    3/2,7/2&    4.8002&    0.132&    1.3054&    0.727 &    &    0.062\\
&    &    &    5/2,7/2&    0.1142&    0.0049&    1.3103&    0.140 &    &    0.012\\
&    &    &    3/2,7/2&    2.2679&    0.0979&    1.4082&    0.626 &    &    0.054\\
$^{27}$Mg&    $^{27}$Al&    0.5,1.5&    1/2,1/2&    70.6176&    0.1122&    0.1122&    0.474 &    4.770&    0.099\\
&    &    &    3/2,1/2&    29.3824&    0.0637&    0.1759&    0.357 &    &    0.075\\
$^{28}$Mg&    $^{28}$Al&    0,2&    1,1&    77.0918&    0.2449&    0.2449&    0.495 &    3.895&    0.127\\
&    &    &    1,1&    22.9082&    0.1541&    0.399&    0.393 &    &    0.101\\
$^{29}$Mg&    $^{29}$Al&    1.5,2.5&    5/2,3/2&    40.8769&    0.0528&    0.0528&    0.460 &    8.709&    0.053\\
&    &    &    1/2,3/2&    4.3222&    0.0126&    0.0654&    0.224 &    &    0.026\\
&    &    &    3/2,3/2&    12.0053&    0.0682&    0.1336&    0.522 &    &    0.060\\
&    &    &    3/2,3/2&    7.1292&    0.0705&    0.2041&    0.531 &    &    0.061\\
&    &    &    5/2,3/2&    13.6818&    0.1784&    0.3825&    0.845 &    &    0.097\\
&    &    &    5/2,3/2&    13.1479&    0.1898&    0.5723&    0.871 &    &    0.100\\
&    &    &    1/2,3/2&    4.1747&    0.0793&    0.6516&    0.563 &    &    0.065\\
&    &    &    3/2,3/2&    2.5205&    0.0683&    0.7199&    0.523 &    &    0.060\\
$^{30}$Mg&    $^{30}$Al&    0,3&    1,2&    85.6711&    0.9115&    0.9115&    0.955 &    4.770&    0.200\\
&    &    &    1,2&    14.3289&    0.4764&    1.3879&    0.690 &    &    0.145\\
$^{31}$Mg&    $^{31}$Al&    0.5,3.5&    1/2,5/2&    29.8291&    0.1792&    0.1792&    0.599 &    7.286&    0.082\\
&    &    &    3/2,5/2&    9.5027&    0.0828&    0.262&    0.407 &    &    0.056\\
&    &    &    1/2,5/2&    19.7349&    0.3625&    0.6245&    0.851 &    &    0.117\\
&    &    &    3/2,5/2&    16.0412&    0.3439&    0.9684&    0.829 &    &    0.114\\
&    &    &    3/2,5/2&    12.988&    0.3282&    1.2966&    0.810 &    &    0.111\\
&    &    &    1/2,5/2&    0.0597&    0.0025&    1.2991&    0.071 &    &    0.010\\
&    &    &    3/2,5/2&    1.1195&    0.0469&    1.346&    0.306 &    &    0.042\\
&    &    &    3/2,5/2&    8.1287&    0.423&    1.769&    0.920 &    &    0.126\\
$^{32}$Mg&    $^{32}$Al&    0,4&    1,3&    99.2926&    1.604&    1.604&    1.266 &    5.508&    0.230\\
&    &    &    1,3&    0.7073&    0.05&    1.6621&    0.224 &    &    0.041\\
&    &    &    1,3&    0.0001&    0.0001&    1.6621&    0.010 &    &    0.002\\
$^{28}$Al&    $^{28}$Si&    3,1&    2,0&    99.6377&    0.0803&    0.0803&    0.750 &    7.286&    0.103\\
&    &    &    4,0&    0.3605&    0.0126&    0.093&    0.297 &    &    0.041\\
&    &    &    3,0&    0.0018&    0.1701&    0.2631&    1.091 &    &    0.150\\
$^{29}$Al&    $^{29}$Si&    2.5,1.5&    3/2,1/2&    90.9415&    0.0896&    0.0896&    0.733 &    8.262&    0.089\\
&    &    &    5/2,1/2&    2.7908&    0.0078&    0.0975&    0.216 &    &    0.026\\
&    &    &    3/2,1/2&    6.1117&    0.0364&    0.1339&    0.467 &    &    0.057\\
$^{30}$Al&    $^{30}$Si&    3,2&    2,1&    8.6243&    0.0118&    0.0118&    0.287 &    10.304&    0.028\\
&    &    &    2,1&    63.6615&    0.1808&    0.1926&    1.125 &    &    0.109\\
&    &    &    3,1&    12.2118&    0.0901&    0.2827&    0.794 &    &    0.077\\
&    &    &    2,1&    7.0753&    0.0523&    0.335&    0.605 &    &    0.059\\
&    &    &    3,1&    4.2795&    0.0389&    0.3739&    0.522 &    &    0.051\\
&    &    &    4,1&    0.2355&    0.0025&    0.3765&    0.132 &    &    0.013\\
&    &    &    4,1&    0.6173&    0.0106&    0.3871&    0.272 &    &    0.026\\
&    &    &    2,1&    1.1&    0.0198&    0.4069&    0.372 &    &    0.036\\
&    &    &    3,1&    0.2221&    0.0117&    0.4186&    0.286 &    &    0.028\\
&    &    &    4,1&    1.2051&    0.0735&    0.4921&    0.717 &    &    0.070\\
$^{31}$Al&    $^{31}$S&    2.5,2.5&    3/2,3/2&    63.5626&    0.1335&    0.1335&    0.895 &    10.666&    0.084\\
&    &    &    5/2,3/2&    12.5331&    0.0732&    0.2067&    0.663 &    &    0.062\\
&    &    &    3/2,3/2&    23.6703&    0.2317&    0.4384&    1.179 &    &    0.111\\
$^{32}$Al&    $^{32}$Si&    1,3&    0,2&    83.1797&    0.2507&    0.2507&    0.867 &    8.262&    0.105\\
&    &    &    2,2&    1.0148&    0.0069&    0.2576&    0.144 &    &    0.017\\
&    &    &    2,2&    3.7196&    0.0732&    0.3309&    0.469 &    &    0.057\\
&    &    &    0,2&    10.5612&    0.3171&    0.648&    0.975 &    &    0.118\\
&    &    &    1,2&    0.2391&    0.0093&    0.6573&    0.167 &    &    0.020\\
&    &    &    1,2&    1.2855&    0.2349&    0.8922&    0.839 &    &    0.102\\
\hline
    \end{tabular}
  \end{center}
\end{table*}

\addtocounter{table}{-1}

\begin{table*}
  \begin{center}
    \leavevmode
    \caption{{\em Continuation.\/}}   

    \label{tab:mgt_comp} 
     \begin{tabular}{lcccccrrrr}
    \hline\hline
Parent &    Daughter&  $J_i, T_i$ & $J_f, T_f$   &    I$_{\beta^-}$&    B(GT) &    Sum B(GT)&    M(GT)&    W&    R(GT)\\ \hline
$^{33}$Al&    $^{33}$Si&    2.5,3.5&    3/2,5/2&    94.8585&    0.5199&    0.5199&    1.766 &    6.746&    0.262\\
&    &    &    7/2,5/2&    0.2942&    0.0122&    0.5321&    0.271 &    &    0.040\\
&    &    &    5/2,5/2&    0.8076&    0.0429&    0.5749&    0.507 &    &    0.075\\
&    &    &    3/2,5/2&    3.2313&    0.1934&    0.7684&    1.077 &    &    0.160\\
$^{31}$Si&    $^{31}$P&    1.5,1.5&    1/2,1/2&    99.4932&    0.019&    0.019&    0.276 &    &    0.041\\
&    &    &    3/2,1/2&    0.5068&    0.0319&    0.0509&    0.357 &    &    0.053\\
$^{32}$Si&    $^{32}$P&    1,2&    1,1&    92.3402&    0.1247&    0.1247&    0.612 &    &    0.091\\
&    &    &    2,1&    7.6598&    0.1232&    0.2479&    0.608 &    &    0.090\\
$^{32}$P&    $^{32}$S&    0,1&    1,0&    100&    0.216&    0.216&    0.465 &    2.754&    0.169\\
$^{33}$P&    $^{33}$S&    1.5,1.5&    3/2,1/2&    100&    0.0174&    0.0174&    0.264 &    6.746&    0.039\\
&    &    &    1/2,1/2&    0&    0.0156&    0.033&    0.250 &    &    0.037\\
&    &    &    5/2,1/2&    0&    0.0864&    1194&    0.588 &    &    0.087\\
$^{34}$P&    $^{34}$S&    1,2&    1,1&    60.2666&    0.0704&    0.0704&    0.460 &    6.746&    0.068\\
&    &    &    1,1&    39.7334&    0.2436&    0.314&    0.494 &    &    0.073\\
$^{35}$P&    $^{35}$S&    0.5,2.5&    3/2,3/2&    7.7626&    0.0038&    0.0038&    0.087 &    6.158&    0.014\\
&    &    &    1/2,3/2&    90.7258&    0.4834&    0.4872&    0.983 &    &    0.160\\
&    &    &    3/2,3/2&    1.5108&    0.1264&    0.6136&    0.503 &    &    0.082\\
$^{36}$Cl&    $^{36}$Ar&    0,1&    1,0&    99.9742&    0.0269&    0.0269&    0.164 &    2.754&    0.060\\
&    &    &    1,0&    0.667&    0.0808&    0.1078&    0.284 &    &    0.103\\
\hline
    \end{tabular}
  \end{center}
\end{table*}

The Gamow-Teller strength  B(GT) is calculated using the following expression,
\begin{equation}
 {B(GT_{\pm})} = \frac{1}{2J_i + 1} f_q^2 \, |{\langle {f}|| \sum_{k}{\sigma^k\tau_{\pm}^k} ||i \rangle}|^2,
\end{equation}
where  $\tau_+|p\rangle = |n\rangle$ , $\tau_-|n\rangle  = |p\rangle$, the index $k$ runs over the single-particle orbitals, 
$|i\rangle$ and $|f \rangle$ describe the state of the parent and daughter nuclei, respectively.
From ref.\cite{mgt} We define
\begin{equation}
M(GT)= [(2j_{i}+1)B(GT)]^{1/2},
\end{equation} 
this is independent of the direction of the transitions. Table I compared calculated and experimental $M(GT)$ values. To get $R(GT)$, we need the total strength, W, and it is defined by
\begin{equation}
  W=\left\{
  \begin{array}{@{}ll@{}}
    |g_{A}/g_{V}|[(2J_{i}+1)3|N_{i}-Z_{i}|]^{1/2} , & for N_{i} \neq  Z_{i},\\
    |g_{A}/g_{V}|[(2J_{f}+1)3|N_{f}-Z_{f}|]^{1/2} , & for N_{i} = Z_{i}. 
  \end{array}\right.
\end{equation} 

Where $R(GT)$ values define as  
\begin{equation}
R(GT) = \frac{M(GT)}{W}.
\end{equation} 

\section{Interaction Hamiltonian}

In our current study, we focus on investigating the properties of $\beta$-decay within the sd-shell model space, utilizing the USDB interaction as our effective Hamiltonian. The USDB Hamiltonian incorporates two-body matrix elements (TBME) and three single-particle energies (SPE) obtained from the renormalized G-matrix. By modifying the G-matrix to reproduce the binding and excitation energies of nuclei with A values ranging from 16 to 40, consistency with experimental data has been taken into account \cite{usdb}.
\par
To comprehensively understand the sd-shell region, \\USDB effective interaction fitted to a total of 608
states across 77 different nuclei. This extensive dataset allows us to explore and analyze various nuclear processes within the sd-shell model space. Specifically, the single-particle energies for the 0d$_{5/2}$, 0d$_{3/2}$, and 1s$_{1/2}$ orbits in the USDB interaction are determined to be 2.117 MeV, -3.9257 MeV, and -3.2079 MeV, respectively.

By employing the USDB interaction, we gain valuable insights into nuclear structure and its implications for $\beta$-decay. Our findings contribute to our knowledge of fundamental nuclear processes and enhance our ability to interpret experimental observations within the sd-shell model space.

\section{Result and Discussion}
\textbf{Comparison of experimental and theoretical $\beta^-$ decay observable}

\subsection{\bf$^{19}$O $\rightarrow$ $^{19}$F }
The experimental half-life of $^{19}$O was determined to be 26.88 msec, which closely matches our calculated value of 26.68 msec\cite{19O}. The $log ft$ values for the nuclear states $5/2_{1}$, $3/2_{1}$, $7/2_{1}$, $5/2_{2}$, and $3/2_{2}$ are 5.801, 4.379, 3.504, 7.72, and 5.003, respectively.

\subsection{\bf$^{20}$O $\rightarrow$ $^{20}$F }
The isotope Oxygen-20 ($^{20}$O) undergoes radioactive decay, with a measured half-life of 13.51 s $\pm$ 0.005 s, transforming into the 1$^{+}$ state \cite{20O}. The decay occurs with branching ratios of 99.973\% and 0.0027\%, respectively. Theoretical calculations predict a half-life of 2.8 s and branching ratios of 97.7005\% and 2.2995\%. Additionally, the calculated $log ft$ values are 3.705 and 3.575, respectively, which closely match the experimental values of 3.3734 and 3.64.
\subsection{\bf$^{21}$O $\rightarrow$ $^{21}$F }
The isotope oxygen-21 ($^{21}$O) initially exists in the 5/2$^+$ state and undergoes radioactive decay with a half-life of 3.42 seconds $\pm$ 0.010. Only the 3/2$^+$ state is distinguishable in the experimental measurements, exhibiting a branching ratio of 37.2\%  and a $log ft$ value of 5.22.

Theoretical calculations have allowed us to determine various states involved in the decay process. These states are denoted as $5/2_{1}$, $3/2_{1}$, $5/2_{2}$, $3/2_{2}$, $7/2_{1}$, $5/2_{3}$, $3/2_{3}$, $7/2_{2}$, and $7/2_{3}$. The calculated branching ratios for these states are 8.0681\%, 33.1716\%, 29.6415\%, 15.4338\%, 1.5963\%, 0.8667\%, 5.0825\%, 6.1384\%, and 0.0012\%, respectively. Additionally, $log ft$ values are determined as 6.361, 5.236, 4.658, 4.915, 5.889, 5.868, 4.991, 4.884, and 8.116 for the corresponding states.
\subsection{\bf$^{22}$O $\rightarrow$ $^{22}$F }
The isotope oxygen-22 ($^{22}$O) undergoes decay with a measured half-life of 2.25 seconds $\pm$ 0.009, while our calculated half-life is 3.58 seconds. The decay of $^{22}$O results in the formation of two states, namely $1_{1}$ and $1_{2}$, with branching ratios of 31\% and 68\%, respectively. According to experimental measurements, the corresponding $log ft$ values for these states are 4.6 and 3.8.

Our theoretical calculations determined a branching ratio of 18.5581\% for the $1_{1}$ state and 81.4419\%for the $1_{2}$ state. The $log ft$ values obtained for these states were 4.578 and 3.607, respectively. 

\subsection{\bf$^{23}$O $\rightarrow$ $^{23}$F }
$^{23}$O undergoes radioactive decay with a measured half-life of 97 ms $\pm$.08 ms. The decay occurs with branching ratios of 99.973\% and 0.0027\% for the respective states. Theoretical calculations predict a different half-life of 50.90 msec for Oxygen-23 decay and branching ratios of 97.7005\%  and 2.2995\% for the two states. Additionally, the calculated $logf t$ values for the decay are 4.497 and 4.337, which closely match the experimental values of ---- and -----.
\subsection{\bf$^{24}$O $\rightarrow$ $^{24}$F }
The isotope Oxygen-24 ($^{24}$O) undergoes a beta-positive decay, transforming into Fluorine-24 ($^{24}$F) in a 1$^+$ state. Initially, it existed in a 0$^+$ state, with an observed lifetime of 72 milliseconds. According to the shell model, the calculated lifetime was 27 milliseconds. The theoretical $log ft$ value for this decay is 3.799, which closely aligns with the experimental value of 4.09.

\subsection{\bf$^{20}$F $\rightarrow$ $^{20}$Ne }
The observed half-life of $^{20}$F is approximately 11.163 seconds with an uncertainty of 0.008 seconds. In theory, we predicted a half-life of 2.242 seconds. Initially, $^{20}$F was in a $2^+$ state and decayed into $^{20}$Ne with a branching ratio of 99.9913\%, which closely aligns with our calculated value of 99.7575\%. The logarithm of the $ft$ value was determined to be 4.816, which is in good agreement with the experimental value of 4.9697.
\begin{figure*}
\begin{center}
\resizebox{1.0\textwidth}{!}{
\includegraphics{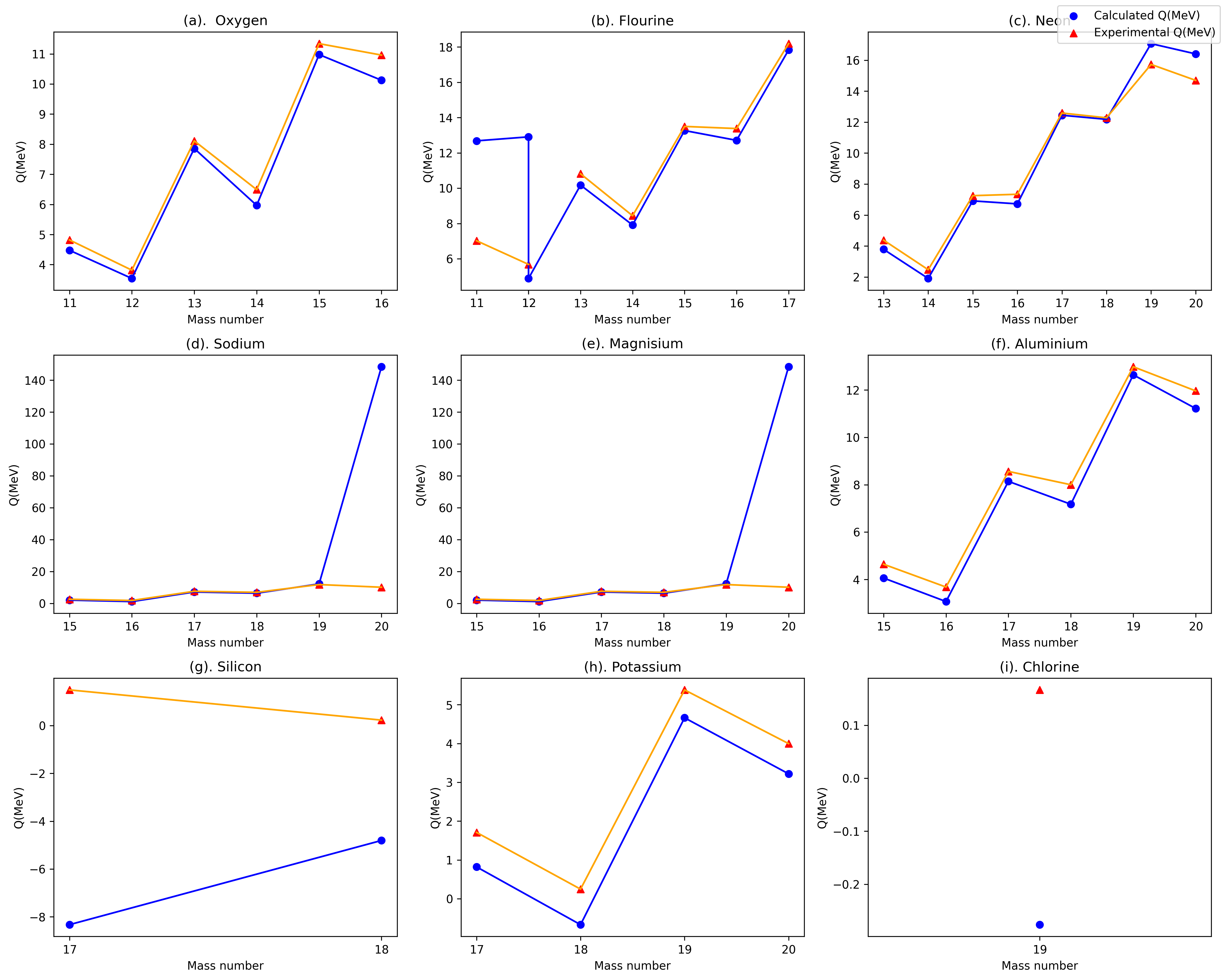} 
}
\end{center}
\caption{
Comparison of experimental and theoretical Q-Value distributions for each nucleus. The x-axis represents the neutron number, and the y-axis represents the Q (MeV) value. Red triangles represent the experimental values, while blue circles represent the theoretical values.}
\label{42Ti_1}
\end{figure*}

\subsection{\bf$^{21}$F $\rightarrow$ $^{21}$Ne }
The decay of $^{21}$F into $^{21}$Ne with a lifetime of 4.158 $\pm$ 0.020 has been observed. The initial state of $^{21}$F was determined to be 5/2$^+$.

Experimental measurements indicate that the branching ratio for the 5/2$^+_1$ state is 74.1\%, while the theoretical calculation yields a value of 71.3677\%. This state's corresponding $log ft$ values are 16.1 (experimental) and 4.65 (theoretical), which closely align with our calculations of 4.585.

Additionally, the branching ratio for the 3/2$^+$ state is found to be 9.6\%, while for the 7/2$^+$ state, it is 16.1\%. These states' associated $log ft$ values are 5.67 (experimental) and 4.72 (theoretical) for the 3/2$^+$ state and 5.427 and 4.605 for the 7/2$^+$ state, respectively.

\subsection{\bf$^{22}$F $\rightarrow$ $^{22}$Ne} 
The isotope oxygen-22 ($^{22}$O), in its excited state at 4$^+$, undergoes radioactive decay into neon-22 ($^{22}$Ne) with a half-life of 4.23 $\pm$ 0.004 seconds. Our calculated half-life for this decay is 0.2522 seconds.

During the decay process, we observe different branching ratios for the excited states of oxygen-22. Specifically, for the 4$^+_1$, 4$^+_2$, and 4$^+_3$ states, the experimental branching ratios are determined to be 3.1\%, 53.9\%, and 7.0\%, respectively. We have calculated branching ratios of 1.2189\%, 52.8062\%, and 3.7067\%, along with $log ft$ values of 6.7, 4.79, and 5.34 for these states (experimental values). Our theoretical calculations yield $log ft$ values of 6.666, 4.606, and 5.53, which are quite close to the measured values.

In addition to decaying into neon-22, oxygen-22 can also undergo decay to states with higher excitation energies, specifically the 5$^+$ and 3$^+$ states. The experimental branching ratios for these decays are determined to be 8.7\% and 16.4\%, respectively. Theoretical calculations predict branching ratios of 18.7559\% and 16.7565\% for these decays, along with $log ft$ values of 4.70, 5.26 (experimental values), and 4.551, 5.082 (theoretical values).

\begin{table*}
 \begin{center}
    \leavevmode
    \caption{Comparison of the theoretical $\beta$-decay half-lives with the experimental data for the concerned nuclei together with the $\beta^-$  probabilities and quenched theoretical sum $B(GT)$ values. }
    \label{tab:mgt_comp} 
    \begin{tabular}{lccccrrrr}
    \hline
        &  &   &   & \multicolumn{2}{c}{$\beta$ decay half-life}  &  \\
      \cline{5-6}
     $^{A}Z_i(J^{\pi})$ & $^{A}Z_f(J^{\pi})$& J$_i$ & T$_i$ & Theoretical  & Experimental & \multicolumn{1}{c}{Sum B(GT)}  & $\beta^-$(\%)  \\  
      \hline
$^{19}$O&$^{19}$F&2.5&1.5&26.68 s&26.88 s $\pm $5&0.0062&100\\
$^{20}$O&$^{20}$F&0&2&2.8 s&13.51 s $\pm $5&0.7674&100\\
$^{21}$O&$^{21}$F&2.5&2.5&3.325 s&3.42 s $\pm $10&0.0017&100\\
$^{22}$O&$^{22}$F&0&3&3.58 s& 2.25 s $\pm $9&0.1027&100\\
$^{23}$O&$^{23}$F&0.5&3.5&50.90 ms&97 ms $\pm $8 &0.1239&100\\
$^{24}$O&$^{24}$F&0&4&27 ms&72 ms $\pm $ 5&0.6174&100\\
$^{20}$F&$^{20}$Ne&2&1&2.242 s&11.163 s $\pm $ 8&0.0594&100\\
$^{21}$F&$^{21}$Ne&2.5&1.5&3.144 s&4.158 s $\pm $20&0.0145&100\\
$^{22}$F&$^{22}$Ne&4&2&.2522 s&4.23 s $\pm $ 4&0.0008&100\\
$^{23}$F&$^{23}$Ne&2.5&2.5&1.275 s&2.23 s $\pm $14&0.0149&100\\
$^{24}$F&$^{24}$Ne&3&3&0.1233 s&382 m$\pm $ 16&0.0111&100\\
$^{25}$F&$^{25}$Ne&2.5&3.5&30 ms&80 ms $\pm $ 9&0.0404&100\\
$^{26}$F&$^{26}$Ne&1&4&4.502 s&9.7 ms $\pm $ 7&0.58732&100\\
&&4&4&6.6 ms&2.2 ms $\pm $1&0.30203&18\\
$^{23}$Ne&$^{23}$Na&2.5&1.5&50.58 s&37.25 s $\pm $10&0.0132&100\\
$^{24}$Ne&$^{24}$Na&0&2&1 m&3.38 m $\pm $2&0.2052&100\\
$^{25}$Ne&$^{25}$Na&0.5&2.5&423 ms&602 ms$\pm $ 8&0.1776&100\\
$^{26}$Ne&$^{26}$Na&0&3&113.9 ms&197 ms $\pm $2&0.776&100\\
$^{27}$Ne&$^{27}$Na&1.5&3.5&22.59 ms&31.5 ms $\pm $13&0.2381&100\\
$^{28}$Ne&$^{28}$Na&0&4&7.624 ms&20 ms $\pm $1&0.8842&100\\
$^{29}$Ne&$^{29}$Na&1.5&4.5&5.828 ms&15 m s$\pm $ 3&0.1173&100\\
$^{30}$Ne&$^{30}$Na&0&5&3.257 ms&7.3 ms $\pm $ 3&0.7586&100\\
$^{24}$Na&$^{24}$Mg&4&0&0.0104 h&14.956 h $\pm $3&0.0079&0.05\\
$^{25}$Na&$^{25}$Mg&2.5&1.5&42.32 s&59.1 s $\pm $6&0.0327&100\\
$^{26}$Na&$^{26}$Mg&3&2&0.2427 s&1.07128 s $\pm $ 25&0.1036&100\\
$^{27}$Na&$^{27}$Mg&2.5&2.5&99.18 ms&301 ms $\pm $6&0.3204&100\\
$^{28}$Na&$^{28}$Mg&1&3&0.0192 s&30.5 s $\pm $4&0.1483&100\\
$^{29}$Na&$^{29}$Mg&1.5&3.5&22.46 ms&44.1 ms $\pm $9&0.0974&100\\
$^{30}$Na&$^{30}$Mg&2&4&7.815 ms&48 m s $\pm $2&0.021&100\\
$^{31}$Na&$^{31}$Mg&1.5&4.5&6.798 ms&17.0 m $\pm $ s&0.0439&100\\
$^{27}$Mg&$^{27}$Al&0.5&1.5&1.137 m&9.458 m $\pm $ 12&0.1122&100\\
$^{28}$Mg&$^{28}$Al&0&2&2.5766 h&20.915 h $\pm $9&0.2449&100\\
$^{29}$Mg&$^{29}$Al&1.5&2.5&0.6615 s&1.30 s $\pm $12&0.0528&100\\
$^{30}$Mg&$^{30}$Al&0&3&119 ms&335 ms $\pm $17&0.9115&100\\
$^{31}$Mg&$^{31}$Al&0.5&3.5&17.55 ms&270 m s $\pm $ 2&0.1792&100\\
$^{32}$Mg&$^{32}$Al&0&4&12.81 ms&86 ms $\pm $ 5&1.604&100\\
$^{28}$Al&$^{28}$Si&3&1&0.1678 m&2.245 m $\pm $2&0.0803&100\\
$^{29}$Al&$^{29}$Si&2.5&1.5&0.38 m&6.56 m $\pm $6&0.0896&100\\
$^{30}$Al&$^{30}$Si&3&2&3.499 s&3.62 s $\pm $ 6&0.0118&100\\
$^{31}$Al&$^{31}$Si&2.5&2.5&313.3 ms&644 m s $\pm $ 25&0.1335&100\\
$^{32}$Al&$^{32}$Si&1&3&21.98 ms&32.3 m s $\pm $3&0.2507&100\\
$^{33}$Al&$^{33}$Si&2.5&3.5&17.25 ms&41.7 ms $\pm $ 2&0.5199&100\\
$^{31}$Si&$^{31}$P&1.5&1.5&9.43 s&157.24 m $\pm $20&0.019&100\\
$^{32}$Si&$^{32}$P&1&2&0.0266 yr&153 yr $\pm $19&0.1247&100\\
$^{32}$P&$^{32}$P&0&1&0.00318 d&14.268 d $\pm $ 5&0.216&100\\
$^{33}$P&$^{33}$P&1.5&1.5&52.96 d&25.35 d $\pm $ 11&0.0174&100\\
$^{34}$P&$^{34}$P&1&2&3.351 s&12.43 s $\pm $10&0.0704&100\\
$^{35}$P&$^{35}$P&0.5&2.5&31.34 s&47.3 s $\pm $8&0.0038&100\\
$^{36}$Cl&$^{36}$Ar&0&1&0.00207 yr&3.01$\times10^5$ yr $\pm $2&0.0269&98.10\\

\hline
 \end{tabular}
   \end{center}
     \end{table*}

\subsection{\bf$^{23}$F$\rightarrow$ $^{23}$Ne }
Initially, $^{23}$F in its 5/2$^+$ state undergoes decay to form $^{23}$Ne with a half-life of 2.23 $\pm$ 0.014 seconds. However, our calculated value for this half-life is 1.275 seconds, indicating a discrepancy. The theoretical branching ratios for the states 5/2$^+_1$, 5/2$^+_2$, and 5/2$^+_3$ are 34.1604, 1.2735, and 27.8797, respectively, while for other excited states, such as 30 and 5.6, the branching ratios have not been resolved yet.

Regarding the 7/2$^+_1$, 7/2$^+_2$, 3/2$^+_1$, and 3/2$^+_2$ states, the theoretical $log ft$ values and branching ratios are 0.6805\%, 1.5773\%, 5.0902\%, and 17.6043\%, respectively. However, the experimental data is inconclusive for the 7/2$^+$ state, as it remains unresolved. On the other hand, for the 3/2$^+_1$ and 3/2$^+_2$ states, the experimental branching ratios are 11 and 12, respectively, with $log ft$ values of 5.6 and 5.0. These experimental results align with the data obtained.

\subsection{\bf$^{24}$F $\rightarrow$ $^{24}$Ne }
$^{24}$F, initially in the 3$^+$ state, undergoes a decay to $^{24}$Ne in the 2$^+$ state with a half-life of 382 ms $\pm$ 0.016. However, our calculated half-life is 123.3 ms. The experimental values for the $log ft$ values and branching ratio are currently unavailable. However, we have theoretically determined a branching ratio of 22.8166\%, 21.922\%, 44.8041\%, 4.4049\% for the states $2_{1}$, $2_{2}$, $4_{1}$, $3_{1}$ respectively. Additionally, the $log ft$ values is calculated as 5.544, 5.293, 4.941, and 5.796 for the respective states. These theoretical calculations can be valuable for conducting experimental studies.

\subsection{\bf$^{25}$F $\rightarrow$ $^{25}$Ne }
$^{25}$F decays with a half-life of 80 ms $\pm$ 0.009. However, we obtained a half-life of 30 ms based on our calculations. The specific decay level of $^{25}$F remains unresolved experimentally. Theoretically, $^{25}$F can decay into several states, namely $3/2_{1}$, $3/2_{2}$, $7/2_{1}$, $7/2_{2}$, $7/2_{3}$, $7/2_{4}$, and $3/2_{2}$, with corresponding branching ratios of 19.6223\%, 3.2086\%, 30.4252\%, 30.4252\%, 0.327\%, 0.327\%, and 2.3347\%. The $log ft$ values for these states are calculated as 4.983, 5.588, 4.507, 4.507, 6.351, 6.351, and 5.463, respectively.
\subsection{\bf$^{26}$F $\rightarrow$ $^{26}$Ne }
The radioactive isotope $^{26}$F undergoes decay into $^{26}$Ne with an experimental half-life of 2.2 ms $\pm$ 0.001. However, our calculation indicates a longer half-life of 6.6 ms. Specifically, $^{26}$F, initially in the $1^+$ state, decays into the $4_{1}$, $3_{1}$, $4_{2}$, $3_{2}$, $5_{1}$, and $5_{2}$ states of $^{26}$Ne. The branching ratios for these decays are determined as 73.6897\%, 3.3661\%, 13.3773\%, 1.9842\%, 1.7723\%, and 5.8103\%, respectively. The corresponding $log ft$ values are calculated as 4.383, 5.499, 4.887, 5.627, 5.43, and 4.853. However, experimentally, the specific decay level of $^{26}$F remains unresolved. These results hold significant potential for guiding future experimental investigations.
\subsection{\bf$^{23}$Ne $\rightarrow$ $^{23}$Na}
The isotope $^{23}$Ne exhibits an experimental half-life of 37.25 seconds $\pm$ 0.010, while our calculation yields a longer half-life of 50.58 seconds. Initially in the 5/2$^+$ state, $^{23}$Ne undergoes decay into the following states: $3/2_{1},1/2$; $5/2_{1},1/2$; $7/2_{1},1/2$; $3/2_{2},1/2$; and $5/2_{2},1/2$. These decays' experimental branching ratios are 67\%, 31.9\%, 1.1\%, and 0.065\%, respectively. Theoretical calculations provide branching ratios of 57.9673\%, 41.2246\%, 0.3014\%, and 0.5064\% for the same states. The experimental $log ft$ values for these states closely align with the theoretical calculations, resulting in values of 5.27, 5.38, 5.82, and 6.13, respectively. The corresponding states' theoretical $log ft$ values are calculated as 5.469, 5.429, 6.44, and 5.685.
\subsection{\bf$^{24}$Ne $\rightarrow$ $^{24}$Na}
The initial state of $^{24}$Ne is $0^+$, and it undergoes decay into the states $1_{1}$ and $1_{2}$ with a lifetime of 3.38 minutes. Theoretical calculations yield a half-life of 1 minute. The experimental branching ratio for the first state is 92.1\%, and for the second state is 7.9\%, while the theoretical branching ratio is measured as 84.8747\% and 15.1253\%, respectively. The measured $log ft$ values for the decay are 4.364 and 4.400, while the calculated values are 4.278 and 4.323, which closely align with the experimental values.
\subsection{\bf$^{25}$Ne $\rightarrow$ $^{25}$Na}
The isotope $^{25}$Ne, initially in the 1/2$^+$ state, undergoes decay with a half-life of 602 ms $\pm$ 0.008. Our calculation yields a slightly shorter half-life of 423 ms. The decay leads to the population of the states $3/2_{1},3/2$; $1/2_{1},3/2$; $3/2_{2},3/2$; $3/2_{3},3/2$; and $1/2_{2},3/2$. These decays' corresponding experimental branching ratios are 76.6\%, 19.5\%, 2.1\%, and 1.2\%, respectively. The $log ft$ values for these states are measured as 4.41, 4.7, 5.26, and 4.82. The theoretical calculations yield a branching ratio of 69.0255\%, 24.4595\%, 4.1392\%, and 1.1953\% for the respective states, with $log ft$ values of 4.341, 4.54, 4.967, and 4.845.
\subsection{\bf$^{26}$Ne $\rightarrow$ $^{26}$Na}
Moving on to the decay of $^{26}$Ne to $^{26}$Na, where the initial state is 0$^+$, the half-life is experimentally determined to be 197 ms $\pm$ 0.002. However, our calculation yields a shorter half-life of 113.9 ms. The experimental branching ratios for the states $1_{1}$, $1_{2}$, and $1_{3}$ are 91.6\%, 4.2\%, and 1.9\%, respectively. The corresponding $log ft$ values are measured as 3.87, 4.8, and 4.7.

Theoretical calculations result in a branching ratio of 83.4445\%, 12.4313\%, and 3.9369\% for the respective states, along with $log ft$ values of 3.7, 4.14, and 4.211.

\subsection{\bf$^{27}$Ne $\rightarrow$ $^{27}$Na}
The isotope $^{27}$Ne, initially in the 3/2$^+$ state, 
undergoes decay with a half-life
of 31.5 ms $\pm$ 0.013. However, our calculation indicates a slightly shorter half-life of 22.59 ms. Experimentally, we have determined the branching ratios and $log ft$ values for the states $5/2_{1},5/2$; $3/2_{1},5/2$; and $1/2_{1},5/2$, which are measured as 59.5\%, 4.2\%, 3.4\%, and 4.4, 5.54, 5.33, respectively. The status of other excited states is still unresolved experimentally. In our theoretical calculations, we obtained $log ft$ values of 4.213, 4.844, 6.092, 4.626, 4.401, 5.467, 5.681, 5.52, and 4.906, along with branching ratios of 64.4845\%, 14.9302\%, 0.4334\%, 6.9833\%, 10.2101\%, 0.6278\%, 0.3812\%, 0.5428\%, and 
\\1.4067\% for the states $5/2_{1},5/2$; $3/2_{1}$; $1/2_{1}$; $5/2_{2}$; $3/2_{2}$; $1/2_{2}$; $5/2_{3}$; $3/2_{3}$; and $1/2_{3}$, respectively. These calculated values are consistent with the existing data and can potentially contribute to resolving the experimental values further.
\subsection{\bf$^{28}$Ne $\rightarrow$ $^{28}$Na}
The isotope $^{28}$Ne undergoes decay with a half-life of 20 ms $\pm$ 0.001. The experimental branching ratios for the decays are determined as 55\%, 1.7\%, 20.1\%, 8.5\%, 1.3\%, and 0.9\%, with the corresponding $log ft$ values of 4.2, 5.3, 4.2, 4.5, 5.2, and 5.3. However, our calculation yields a shorter half-life of 7.624 ms and different branching ratios: 72.7406\%, 14.7253\%, 1.9715\%, 2.9957\%, 7.4902\%, and 0.0766\%, respectively. The calculated $log ft$ values for the states $1_{1}$, $1_{2}$, $1_{3}$, $1_{4}$, $1_{5}$, and $1_{6}$ are 3.643, 3.967, 4.725, 4.444, 3.759, and 5.716.
\subsection{\bf$^{29}$Ne $\rightarrow$ $^{29}$Na}
Experimental measurements indicate that $^{29}$Ne decays with a half-life of 15 ms $\pm$ 0.003. However, our calculation yields a slightly shorter half-life of 5.828 ms. The specific decay levels of $^{29}$Ne are still undetermined, but we have an existing value of the branching ratio of 33\% for the $5/2^+$ level. In our calculations, we obtained branching ratios of 24.0716\%, 24.0716\%, 1.5023\%, 21.6126\%, 21.6126\%, 3.394\%, 5.0929\%, 0.8485\%, and 0.8485\% for the states $5/2_{1},7/2$; $5/2_{2}$; $1/2_{1}$; $5/2_{3}$; $5/2_{4}$; $1/2_{2}$; $1/2_{3}$; $5/2_{5}$; and $5/2_{6}$, respectively. Additionally, we calculated $log ft$ values of 4.521, 4.521, 5.419, 4.068, 4.068, 5.852, 4.588, 5.344, and 5.344 for these respective states.
\subsection{\bf$^{30}$Ne $\rightarrow$ $^{30}$Na}
Initially, $^{30}$ was in the 0$^+$ state and decayed with a half-life of 7.3 ms $\pm$ 0.003 into the states $1_{1}$, $1_{2}$, and $1_{3}$. The experimental branching ratios for these states are measured as 63\%, 7.7\%, and 14\%, respectively. In our calculations, we obtained branching ratios of 60.3304\%, 23.1876\%, and 16.482\% for the states $1_{1}$, $1_{2}$, and $1_{3}$, respectively. Additionally, the calculated logarithm of the $log ft$ values for these decays is 3.71, 3.731, and 3.693, which are comparable to the experimental values.

\subsection{\bf$^{24}$Na $\rightarrow$ $^{24}$Mg}
The isotope $^{24}$, initially in the 4$^+$ state, undergoes decay into the states $4_{1}$ and $3_{1}$ with a branching ratio of 99.867\% and 0.070\%, respectively. The $log ft$ values for these decays are measured as 6.12 and 6.66. In our calculations, we obtained a branching ratio of 76.2805\% for the $4_{1}$ state and 19.7423\% for the $3_{1}$ state, while the corresponding $log ft$ values are calculated as 5.693 and 6.011. These calculated values are in agreement with the experimental values. The experimental half-life of the decay is determined as 14.956 hours $\pm$ 0.003, while our calculated value is 0.0104 hours.
\subsection{\bf$^{24}$Na $\rightarrow$ $^{24}$Mg}
The isotope $^{24}$Na, initially in the 0$^+$ state, has a half-life of 3.38 minutes $\pm$ 0.002. However, our calculation yields a significantly shorter half-life of 1 minute. The experimental branching ratios for the decays are measured as 92.1\% and 7.9\% for the states $1_{1}$ and $1_{2}$, respectively. The corresponding $log ft$ values are determined as 4.364 and 4.4.

Theoretical calculations result in branching ratios of 84.8747\% and 15.1253\% for the respective states, with $log ft$ values of 4.278 and 4.323.

\subsection{\bf$^{25}$Na $\rightarrow$ $^{25}$Mg}
Initially, $^{25}$Na was in the 5/2$^+$ state and underwent decay into states $5/2_{1}$, $3/2_{1}$, $7/2_{1}$, $5/2_{2}$, and $3/2_{2}$. These decays' experimental branching ratios are 68.0102\%, 24.5795\%, 6.6518\%, 0.3719\%, and 0.385\%. The corresponding $log ft$ values for these states are measured as 5.075, 4.869, 4.953, 5.95, and 4.899. Our theoretical calculations obtained branching ratios of 68.0102\%, 24.5795\%, 6.6518\%, 0.3719\%, and 0.385\%. The calculated $log ft$ values are 5.075, 4.869, 4.953, 5.95, and 4.899. These calculated values are quite comparable to the experimental values.
\subsection{\bf$^{26}$Na $\rightarrow$ $^{26}$Mg}
Initially, $^{26}$Na was in the $3^+$ state and underwent decay with a half-life of 1.01728 s $\pm$ 0.025. The experimental branching ratios for the decay into different states are measured as 87.8\%, 0.05\%, 1.31\%, 3.17\%, 0.493\%, 1.65\%, 2.378\%, 4.71\%, 0.0129\%, and 1.72\%. These states' corresponding $log ft$ values are 7.6, 5.87, 5.33, 6.15, 5.62, 5.25, 7.31, and 4.74. In our theoretical calculations, we obtained branching ratios of 76.1699\%, 0.5895\%, 1.108\%, 8.2572\%, 0.8785\%, 2.5027\%, 3.3914\%, 0.0186\%, and 4.2126\% for these states. The calculated $log ft$ values are also 4.575, 6.427, 5.924, 4.929, 5.888, 5.408, 5.143, and 7.239. These calculated values closely align with the experimental values.
\subsection{\bf$^{27}$Na $\rightarrow$ $^{27}$Mg}
$^{27}$Na, initially in the 5/2$^+$ state, undergoes decay with a measured half-life of 301 ms $\pm$ 0.006. It decays into the states $3/2_{1}$, $5/2_{1}$, $5/2_{2}$, $7/2_{1}$, $3/2_{2}$, $7/2_{2}$, and $3/2_{3}$. The experimental branching ratios for these decays are determined as 85.8\%, 11.3\%, 0.5\%, 0.5\%, 0.52\%, 0.74\%, and 0.026\%. The corresponding $log ft$ values for these states is measured as 4.3, 4.99, 6.3, 5.91, 5.76, 5.63, and 6.81. Some higher excited states are still unresolved experimentally.

In our theoretical calculations, we obtained branching ratios of 83.1719\%, 6.8371\%, 1.3168\%, 0.6022\%, 0.3861\%, 0.956\%, 0.8395\%, 0.3604\%, 0.7471\%, 0.5299\%, 1.6929\%, and 0.0019\%. Additionally, the calculated $log ft$ values is 4.084, 5.01, 5.668, 5.69, 5.771, 5.376, 5.39, 5.591, 5.187, 5.21, 4.62, and 7.505 for the states $3/2_{1}$, $5/2_{1}$, $5/2_{2}$, $7/2_{1}$, $3/2_{2}$, $7/2_{2}$, $3/2_{3}$, $5/2_{3}$, $5/2_{4}$, $7/2_{2}$, and $5/2_{5}$, respectively.

These calculated values closely match the experimental values and provide important information for further understanding the decay of $^{27}$Na.

\subsection{\bf$^{28}$Na $\rightarrow$ $^{28}$Mg}
$^{28}$Na, initially in the 1$^+$ state, undergoes decay to various states including $0_{1}$, $2_{1}$, $0_{2}$, $2_{2}$, $1_{1}$, $2_{3}$, $1_{2}$, $2_{4}$, $2_{5}$, $0_{3}$, $2_{6}$, and $1_{3}$. The experimental branching ratios for these decays are determined as 60.4\%, 11\%, 20.1\%, 1\%, 3.2\%, 0.2\%, 1.5\%, $\leq$0.1\%, 0.3\%, 0.2\%, 0.5\%, and 1.4\%. The corresponding $log ft$ values for these states are measured as 4.6, 5.1, 4.42, 5.6, 5.1, 6.2, 5.2, $>$6.4, 5.8, 5.8, 5.2, and 4.7. Some other decay levels are still unresolved experimentally.

In our theoretical calculations, we obtained branching ratios of 59.6477\%, 4.2164\%, 21.1598\%, 2.3639\%, 3.5429\%, 0.0136\%, 2.1012\%, 0.1946\%, 1.3337\%, 0.3256\%, 0.2251\%, 0.6703\%, 0.4852\%, 0.139\%, and 0.1039\% for the states $0_{1}$, $2_{1}$, $0_{2}$, $2_{2}$, $1_{1}$, $2_{3}$, $1_{2}$, $2_{4}$, $2_{5}$, $0_{3}$, $2_{6}$, $1_{3}$, $0_{4}$, $1_{4}$, $2_{7}$, and $2_{8}$, respectively. Additionally, the calculated $log ft$ values are 4.419, 5.332, 4.172, 5.011, 4.809, 7.197, 4.84, 5.862, 4.9, 5.375, 5.435, 4.931, 5.05, 5.49, 5.576, and 4.136.
These calculated values closely match the experimental values, providing valuable insights into the decay of $^{28}$Na. Furthermore, they can be utilized to study the unresolved decay levels in more detail.
\subsection{\bf$^{29}$Na $\rightarrow$ $^{29}$Mg}
$^{29}$Na, initially in the 3/2$^+$ state, undergoes decay with an experimental half-life of 44.1 ms $\pm$ 0.009. However, our theoretical calculations yield a slightly different half-life of 22.46 ms. Currently, only the decay level of 3/2$^+$ has been resolved experimentally.
Nevertheless, we have theoretically evaluated the branching ratios for the decays to $3/2_{1}$, $1/2_{1}$, $5/2_{2}$, $3/2_{2}$, $1/2_{2}$, $5/2_{3}$, $3/2_{3}$, $5/2_{4}$, $5/2_{5}$, and $3/2_{4}$ as 35.1635\%, 11.581\%, 0.1664\%, 0.5719\%, 28.3018\%, 4.4167\%, 10.6317\%, 3.8083\%, 0.3995\%, and 4.8335\%, respectively. Additionally, the corresponding $log ft$ values for these states are determined as 4.602, 5.077, 6.661, 5.991, 4.239, 4.942, 4.488, 4.92, 5.746, and 7.8. These theoretical values provide valuable insights into the decay process of $^{29}$Na and can be used to study the decay levels that are currently unresolved experimentally.
\subsection{\bf$^{30}$Na $\rightarrow$ $^{30}$Mg}
$^{30}$Na, initially in the 2$^+$ state, undergoes decay to various states including $2_{1}$, $2_{2}$, $3_{1}$, $2_{3}$, $2_{4}$, $1_{1}$, $1_{2}$, $3_{2}$, $2_{5}$, $2_{6}$, $3_{3}$, $2_{7}$, $3_{4}$, $1_{3}$, $2_{4}$, $2_{5}$, $3_{5}$, $3_{6}$, $1_{4}$, and $3_{7}$. The corresponding $log ft$ values for these states is 5.269, 4.996, 6.018, 4.371, 5.674, 4.892, 4.736, 4.772, 4.726, 6.602, 4.75, 5.612, 5.336, 4.87, 5.701, 4.823, 7.244, 4.769, 6.018, and 4.851, while the branching ratios for these decays are 9.4971\%, 10.3898\%, 0.6652\%, 28.2814\%, 1.2435\%, 7.4875\%, 8.878\%, 5.8731\%, 6.0893\%, 0.0782\%, 5.1763\%, 0.6439\%, 1.1474\%, 3.2149\%, 0.4182\%, 2.7423\%, 0.0097\%, 2.5133\%, 0.1351\%, and 1.793\%.

Experimentally, the branching ratios for the decay to some of these states are 9.5\% and 3.8\%, with the corresponding $log ft$ values of 5.86 and 6.12. However, the excitement level is still unresolved experimentally.

These findings provide valuable information about the decay process of $^{30}$Na, but further experimental investigations are needed to resolve the excited level and validate the theoretical predictions.
\subsection{\bf$^{31}$Na $\rightarrow$ $^{31}$Mg}
The radioactive isotope $^{31}$Na undergoes decay with a half-life of 17.0 minutes $\pm$ 0.004 seconds. However, our theoretical calculation yields a half-life of 6.798 milliseconds, indicating a significant discrepancy. Experimentally, the following states have been observed: $3/2_{1}$, $5/2_{1}$, $1/2_{1}$, $5/2_{2}$, $3/2_{2}$, and $1/2_{2}$. The corresponding branching ratios for these states are 8\%, 0.68\%, 22\%, 1.04\%, 5.4\%, and 21.8\%, respectively. The $log ft$ values for these states are 5.4, 6.4, 5, 6, 5.5, and 4.7, respectively.

Theoretical calculations result in branching ratios of 9.69\%, 9.365\%, 9.5712\%, 24.8454\%, 14.5603\%, and 16.7288\%, along with $log ft$ values of 4.948, 4.765, 4.604, 4.082, 4.268, and 4.106, for the aforementioned states. Additionally, we further calculate a branching ratio of 5.7916\%, 4.8002\%, 0.1142\%, and 2.2679\%, with $log ft$ values of 4.48, 4.469, 5.901, and 4.599, for the states $5/2_{3}$, $3/2_{3}$, $5/2_{4}$, and $3/2_{4}$, respectively. However, experimental data for these states is currently unavailable. 

\subsection{\bf$^{27}$Mg $\rightarrow$ $^{27}$Al}
$^{27}$Mg decays with a half-life of 9.458 minutes $\pm$ 0.012, while our calculated half-life is 1.137 minutes. The decay leads to the states $1/2_{1}$ and $3/2_{1}$ with a branching ratio of 70.94\% and 29.06\% respectively. The $log ft$ values for these decays are 4.7297 and 4.934, respectively. Theoretical calculations yield a branching ratio of 70.6176\% and 29.3824\%, along with a $log ft$ values of 4.54 and 4.786, which closely match the experimental values.

\subsection{\bf$^{28}$Mg $\rightarrow$ $^{28}$Al}
$^{28}$Mg undergoes decay with a half-life of 20.915 hours $\pm$ 0.009, originating from the initial 0$^+$ state and transitioning to states $1_{1}$ and $1_{2}$. The theoretical half-life for this decay process is calculated to be 2.5766 hours. The branching ratio for the decay is determined to be 94.8\% and 4.9\% for the respective states, with $log ft$ values of 4.453 and 4.57. Our calculations result in a branching ratio of 77.0918\% and 22.9082\%, with a $log ft$ values of 4.201 and 4.402, which differ slightly from the experimental values.

\subsection{\bf$^{29}$Mg $\rightarrow$ $^{29}$Al}
$^{29}$Mg (3/2$^+$) exhibits decay with a half-life of 1.30 seconds $\pm$ 0.012, while our calculations determine a half-life of 0.6615 seconds. The experimental branching ratios for the states $5/2_{1}$, $1/2_{1}$, $3/2_{1}$, $3/2_{2}$, $5/2_{2}$, $5/2_{3}$, $1/2_{2}$, $3/2_{3}$, $5/2_{3}$, $3/2_{3}$, $5/2_{4}$, and $1/2_{3}$ are measured as 27\%, 7\%, 21\%, 7.8\%, 6\%, 28\%, 3\%, 0.9\%, 0.35\%, 0.3\%, 0.2\%, and $\leq$1\%, respectively, with the corresponding $log ft$ values of 5.32, 5.49, 4.73, 4.9, 4.93, 4.21, 5.06, 5.5, 5.9, 5.8, 5.8, >5.2, and >5.8. The theoretical calculations yield branching ratios of 40.8769\%, 4.3222\%, 12.0053\%, 7.1292\%, 13.6818\%, 13.1479\%, 4.1747\%, 2.5205\%, 0.3517\%, 0.4614\%, 0.0016\%, 0.1354\%, and 0.0026\%, along with $log ft$ values of 4.867, 5.491, 4.756, 4.742, 4.339, 4.312, 4.691, 4.756, 5.498, 4.794, 7.706, and 5.677. These theoretical values align well with the experimental measurements.

\subsection{\bf$^{30}$Mg $\rightarrow$ $^{30}$Al}
$^{30}$Mg, initially in the 0$^+$ state, undergoes decay to the states $1_{1}$ and $1_{2}$ with measured branching ratios of 68\% and 7\% respectively. The $log ft$ values for these states is measured as 3.96 and 4.3. Our calculations yield a branching ratio of 85.6711\% and 14.3289\%, along with $log ft$ values of 3.63 and 3.912 for the corresponding states.

\subsection{\bf$^{31}$Mg $\rightarrow$ $^{31}$Al}
$^{31}$Mg, which exhibits a 1/2$^+$ state, undergoes decay with a half-life of 270 milliseconds $\pm$ 0.002. Theoretical calculations predict a half-life of 17.55 milliseconds. The decay levels of $^{31}$Mg remain unresolved experimentally. Theoretically, we calculate the branching ratios for the states $1/2_{1}$, $3/2_{1}$, $1/2_{2}$, $3/2_{2}$, $3/2_{3}$, $1/2_{3}$, $3/2_{4}$, $3/2_{5}$, $3/2_{6}$, and $1/2_{4}$ to be 29.8291\%, 9.5027\%, 19.7349\%, 16.0412\%, 12.988\%, 0.0597\%, 1.1195\%, 8.1287\%, 0.1964\%, 0.5841\%, and 0.1885\%, respectively. The corresponding $log ft$ values for these states is 4.337, 4.672, 4.031, 4.054, 4.074, 6.199, 4.919, 3.964, 5.417, and 4.936 for $1/2_{1}$, $3/2_{1}$, $1/2_{2}$, $3/2_{2}$, $3/2_{3}$, $1/2_{3}$, $3/2_{4}$, $3/2_{5}$, $3/2_{6}$, and $1/2_{4}$, respectively.

\subsection{\bf$^{32}$Mg $\rightarrow$ $^{32}$Al}
Initially, $^{32}$Mg was in the 0$^+$ state and undergoes decay to the states $1_{1}$, $1_{2}$, and $1_{3}$. The measured half-life for this decay is 86 milliseconds $\pm$ 0.005, while our calculations predict a half-life of 12.81 milliseconds. The measured branching ratios for the decay to the states $1_{1}$, $1_{2}$, and $1_{3}$ are 55\%, 24.6\%, and 10.7\% respectively, with the corresponding $log ft$ values of 4.4, 4.1, and 4.4. The calculated branching ratio for the same states is 99.2926\%, 0.7073\%, and 0.0001\%, respectively, with corresponding $log ft$ values of 3.385, 4.826, and 7.824.

\subsection{\bf$^{28}$Al $\rightarrow$ $^{28}$Si}
$^{28}$Al undergoes decay into the state $2_{1}$ with a measured branching ratio of 99.99\% and a $log ft$ value of 4.8664. Our calculations yield a calculated branching ratio of 99.6377\% and a $log ft$ value of 4.605. The measured half-life for this decay is 2.245 minutes $\pm$ 0.002, while the calculated half-life is 0.1678 minutes.
\subsection{\bf$^{29}$Al $\rightarrow$ $^{29}$Si}
$^{29}$Al undergoes decay with a half-life of 6.56 minutes $\pm$ 0.006, while our calculations predict a half-life of 0.38 minutes. The branching ratios for the states $3/2_{1}$, $5/2_{1}$, $3/2_{2}$, and $5/2_{2}$ are measured to be 89.9\%, 3.8\%, 6.3\%, and 0.033\% respectively. The corresponding $log ft$ values for these states are 5.05, 5.733, 5.026, and 6.11. Theoretical calculations result in branching ratios of 90.9415\%, 2.7908\%, 6.1117\%, and 0.1558\% for the respective states, with the $log ft$ values of 4.638, 5.695, 5.029, and 5.863.

\subsection{\bf$^{30}$Al $\rightarrow$ $^{30}$Si}
$^{30}$Al decays with a half-life of 3.62 seconds $\pm$ 0.006, while the theoretical value is 3.499 seconds. The measured branching ratios for the states $2_{1}$, $2_{2}$, $3_{1}$, $2_{3}$, $3_{2}$, and $4_{1}$ are 17.1\%, 67.3\%, 5.7\%, 6.6\%, 2.6\%, and 0.16\% respectively. The corresponding $log ft$ values for these states are 5.619, 4.578, 5.06, 4.985, 5.17, and 5.92. The calculated branching ratios for the same states are 8.6243\%, 63.6615\%, 12.2118\%, 7.0753\%, 4.2795\%, and 0.2355\% respectively, with the $log ft$ values of 5.52, 4.333, 4.635, 4.871, 5, and 6.186. The calculated values closely match the experimental measurements.
\subsection{\bf$^{31}$Al $\rightarrow$ $^{31}$Si}
$^{31}$Al, in a state of 5/2$^+$, decays with a measured lifetime of 644 milliseconds $\pm$ 0.025, while the calculated lifetime is 3.499 seconds. The experimental decay levels are still unresolved, but we have calculated the branching ratios for the states to be 63.5626\%, 12.5331\%, 23.6703\%, 0.1089\%, 0.0974\%, and 0.0277\%. The corresponding $log ft$ values for these states are 4.465, 4.725, 4.225, 6.36, 6.014, and 6.075. These values can be utilized for further experimental investigations.
\subsection{\bf$^{32}$Al $\rightarrow$ $^{32}$Si}
$^{32}$Al undergoes decay with a measured half-life of 32.3 milliseconds $\pm$ 0.003, while our calculated half-life is 21.98 milliseconds. The decay process includes branching ratios of 85\%, 4.7\%, 3\%, 4.3\%, and 1.7\% for the corresponding states $0_{1}$, $2_{1}$, $2_{2}$, $0_{2}$, and $1_{1}$. The $log ft$ values for these states is 4.35, 5.3, 5, 4.6, and 4.8, respectively. Theoretical calculations yield a branching ratio of 83.1797\%, 1.0148\%, 3.7196\%, 10.5612\%, and 0.2391\% for the states $0_{1}$, $2_{1}$, $2_{2}$, $0_{2}$, and $1_{1}$, with the $log ft$ values of 4.191, 5.748, 4.725, 4.089, and 5.62, respectively.
\subsection{\bf$^{33}$Al $\rightarrow$ $^{33}$Si}
$^{33}$Al in its 5/2$^+$ state decays with a measured half-life of 41.7 milliseconds and a calculated half-life of 17.25 milliseconds. The experimental decay levels for $^{33}$Al are still unresolved. However, we have theoretically determined the branching ratios for the states $3/2_{1}$, $7/2_{1}$, $5/2_{1}$, $3/2_{2}$, and $5/2_{2}$ to be 94.8585\%, 0.2942\%, 0.8076\%, 3.2313\%, and 0.1936\%, respectively. The $log ft$ values for these states is 3.874, 5.504, 4.958, 4.304, and 5.119, respectively.

\subsection{\bf$^{31}$Si $\rightarrow$ $^{31}$P}
$^{31}$Si in its 3/2$^+$ state undergoes radioactive decay with a measured half-life of 157.25 minutes. The theoretical calculation predicts a half-life of 9.43 seconds. The measured branching ratio for the states $1/2_{1}$ and $3/2_{1}$ is 99.9446\% and 0.0554\% respectively, with the corresponding $log ft$ values of 5.525 and 5.747. These experimental findings are consistent with our theoretical calculation, which yields a branching ratio of 99.4932\% for the $1/2_{1}$ state and 0.5068\% for the $3/2_{1}$ state, along with $log ft$ values of 5.312 and 5.086 respectively.

\subsection{\bf$^{32}$Si $\rightarrow$ $^{32}$P}
$^{32}$Si undergoes radioactive decay with a measured half-life of 153 years $\pm$ 0.019, while our calculations indicate a much shorter half-life of 0.0266 years. It initially exists in a 0$^+$ state and eventually decays into a 1$^+$ state with a branching ratio of 100\% and a corresponding $log ft$ value of 8.21. However, our theoretical calculations suggest a slightly lower branching ratio of 92.3402\%  and a $log ft$ value of 4.494, which deviates from the experimental measurement.

\subsection{\bf$^{32}$P $\rightarrow$ $^{32}$S}
$^{32}$P undergoes radioactive decay with a measured half-life of 14.268 days $\pm$ 0.005, while our calculations indicate a much shorter half-life of 0.00318 days. It initially exists in a 0$^+$ state and decays to a 1$^+$ state with a measured branching ratio of 100\% and a corresponding $log ft$ value of 7.9002. Our calculations align with the experimental measurement, showing a branching ratio of 100\% and a $log ft$ value of 4.256.
\subsection{\bf$^{33}$P $\rightarrow$ $^{33}$S}
$^{33}$P, initially in the 1/2$^+$ state, undergoes decay to the 3/2$^+$ state with a measured half-life of 25.35 days $\pm$ 0.011. In our calculations, we obtained a slightly longer half-life of 52.96 days. The measured branching ratio for this decay is 100\%, accompanied by a log ft value of 5.022, which closely matches our calculated values of a branching ratio of 100\% and a log ft value of 5.35.
\subsection{\bf$^{35}$P $\rightarrow$ $^{35}$S}
$^{35}$P, in the 1/2$^+$ state, undergoes decay to states $3/2_{1}$, $1/2_{1}$, and $3/2_{2}$. The measured half-life for this decay is 47.3 seconds $\pm$ 0.008, while our calculations yield a slightly shorter half-life of 31.34 seconds. The measured branching ratios for the decay are $\leq$0.7\%, 98.8\%, and 0.47\%, accompanied by a $log ft$ value of $>$7.3, 4.122, and 4.96, respectively. Our calculated branching ratios are 7.7626\%, 90.7258\%, and 1.5108\%, with corresponding $log ft$ values of 6.012, 3.906, and 4.488.\par 


Figure 1 showcases the remarkable alignment between the calculated half-lives of nuclei (Z = 8-17) and the corresponding experimental data. The close correspondence highlights the accuracy of the theoretical predictions.

Equally impressive, Figure 2 vividly portrays the Q-values of diverse nuclei, which strikingly resemble the currently available experimental data. This alignment underscores the reliability of the theoretical model in capturing the intricate energy dynamics within nuclei.

Table 1 provides a detailed comparison between each decay state's theoretical and experimental levels. The calculated results derived from the shell model using the USDB interaction are also presented. The experimental data used in this analysis is referenced from \cite{nndc}. Moreover, the table displays the theoretical and experimental values for the branching ratio in columns 5 and 6 and the logarithm of the ft value in columns 7 and 8.

In Table 2, we present a comprehensive overview of the decay process for each parent nucleus in the top row, revealing their transformation into daughter nuclei in the second row. The third and fourth rows provide insights into the initial and final angular momentum isospin associated with the respective nuclear states.

Continuing in Table 2, the fifth row uncovers the branching ratios, shedding light on the likelihood of specific decay pathways. Additionally, the sixth row quantifies each transition's cumulative B(GT) strengths, further illuminating the underlying nuclear structure.

Lastly, the seventh row in Table 2 showcases the M(GT) values, while the eighth and ninth rows present the W and R(GT) values for each transition, providing valuable information about the respective energy redistribution and nuclear response characteristics. This comprehensive table offers a detailed perspective on the intricate decay processes and their associated properties. 

Table 3 presents the decay characteristics of various nuclei, with their corresponding theoretical and experimental half-lives listed in the fourth and fifth columns, respectively. The final row of the table displays the $\beta^-$ probabilities associated with each decay case, providing insights into the likelihood of this specific decay mechanism.
\section{Conclusions }

In this study, we have extensively investigated the properties of nuclei within the range of Z = 8-20 and 8 $\geq$ N $\leq$ 20, employing the USDB interaction in conjunction with the nuclear shell model. Our calculations encompassed various parameters, including half-lives, $log ft$ (log ft) values, Q values, branching ratios, B(GT) values, and M(GT) values. Specifically, we focused on the $0d_{3/2}$, $0d_{5/2}$, and $1s_{1/2}$ shells.

By analyzing a total of 47 $\beta^-$ nuclei, we compared our calculated values with the corresponding experimental data. Remarkably, our findings consistently aligned with the experimental results, reinforcing the credibility of our calculations. The comparison encompassed several crucial aspects such as half-lives, Q values, energy levels, branching ratios, and log ft values. In each of these areas, our computed values exhibited remarkable agreement with the existing experimental data.

Moreover, our investigation extended beyond the realms of available experimental data. Our calculations hold substantial significance for the nuclei where experimental information was scarce. In this regard, our results provide valuable insights that can serve as a foundation for further studies on these nuclei.

Our comprehensive study utilizing the USDB interaction within the nuclear shell model framework has yielded significant achievements. The agreement between our calculations and experimental data, along with the exploration of nuclei with limited experimental information, underscores the importance and applicability of our findings.

\vspace{2mm}

{\bf{Acknowledgments}}\\
S Siwach would like to acknowledge the contributions of individuals who provided valuable insights and assistance throughout the research process.


\end{document}